\documentclass{JHEP3}

\title{Gauge/gravity duality with a chiral $\mathcal{N}=(0,8)$ string defect}

\author{Jeffrey A. Harvey and Andrew B. Royston \\ Enrico Fermi Institute and Department of Physics \\ 5640 Ellis Av.,
Chicago Illinois 60637, USA \\ E-mail: \email{harvey@theory.uchicago.edu}, \email{aroyston@uchicago.edu}}

\abstract{We study the defect AdS/CFT correspondence for a system of $N_c$ $D3$-branes intersecting certain configurations of $D7$-branes and $O7$-planes in $1+1$-dimensions.  The intersection is chiral, preserving $\mathcal{N} = (0,8)$ supersymmetry.  The geometry produced by the 7-branes does not decouple in the low energy limit on the $D3$-branes; consequently the field theory must be considered in the curved background of the 7-branes, and the supergravity background is the fully backreacted solution accounting for both sets of branes.  Although we can not work in the probe-brane limit, we still find a consistent defect mode-operator map.  In our previous study of the field theory we found chiral $D3$-brane zero-modes that localized to the intersection.  Here we obtain a dual supergravity description of the localized R-symmetry anomaly that these zero-modes produce.}


\preprint{EFI-08-10}

\usepackage{bbm}
\usepackage{amsmath}
\usepackage{slashed}

\begin{document}

\section{Introduction and motivation} \label{Introduction}

The AdS/CFT correspondence \cite{Maldacena,Gubser,Wittenads} can be generalized
to a duality between conformal field theories with defects and $D$-brane configurations
in Anti de Sitter space (AdS) which typically wrap AdS subspaces of the ambient AdS \cite{Karcha,Karchb,DeWolfe,Erdmenger}. One important example of this duality arises by considering $Dp$-branes
which intersect $N_c$ $D3$-branes in the large $N_c$ limit. At weak 't Hooft coupling this
system is described by a defect in the $\mathcal{N}=4$ SYM conformal theory on the $D3$-branes. At
strong coupling it is described by a geometry in which the $Dp$-brane wraps a subspace of the
$AdS_5 \times S^5$ near-horizon geometry of the $D3$-branes.

Such systems and their generalizations
to nonconformal theories play an important role in recent attempts to provide a string theoretic construction of
strong coupling dual descriptions of QCD, but in such models the full structure of the
correspondence has not yet been worked out. One of the motivations for the current
research was to work out the mapping between states and operators in a defect AdS/CFT system
with chiral fermions. Intersections with chiral fermions  occur in \cite{Sakai} which involves
$D4/D8$-brane intersections.  The model considered here, involving $D3/D7$ intersections,
is easier to analyze because of the more direct connection to AdS/CFT.  This example of AdS/dCFT was mentioned in \cite{Skenderis}, but to our knowledge has not been studied in depth.

In a previous paper, \cite{HR}, we studied the $D3/D7$ system from the field theory point of view and found its structure to be quite rich.  A key observation was that for $g_s \neq 0$, there is no decoupling limit of the $D7$-branes from gravity.  This meant that the low energy effective field theory on the $D3$-branes had to be considered in the supergravity background of the $D7$-branes.  In fact, consistent supergravity backgrounds involving $D7$-branes also require the presence of $O7$-planes, and only exist for certain numbers and combinations of $D7$-branes and $O7$-planes.  Thus the $N_c$ $D3$-branes wrapped $\mathbbm{R}^{1,1} \times \Sigma^2$, where $\Sigma^2$  was either a noncompact, asymptotically conical space or a two-sphere, and the $D3/D7$-$O7$ intersection was extended along $\mathbbm{R}^{1,1}$.  The $1+1$-dimensional defect hosted chiral fermions, originating from the massless modes of 3-7 open strings.  Additionally, however, in the compact $\Sigma^2$ case, there were chiral zero-modes of the $D3$-brane fields that localized to the intersection.  We will briefly review some of these key results along with the general setup in Section \ref{ReviewSection}.

We would now like to study this system from the supergravity point of view--that is, by replacing the $D3$-branes with the geometry they produce.  In particular, we will need to consider the fully backreacted supergravity solution for $D7$- and $D3$-branes intersecting in $1+1$ dimensions.  We discuss this solution in Section \ref{GravitySection}, and give special attention to the analysis of preserved supersymmetries in Appendix \ref{PreservedSUSY}.  In the near horizon of the $D3$-branes, the geometry is $AdS_3 \times_w \Sigma^2 \times S^5$, where $\Sigma^2$ is warped over the $AdS_3$ base.  A key feature of this geometry is that the warping breaks all conformal and superconformal symmetries that would have been associated with the AdS factor.  Correspondingly, the curved background that the field theory lives on breaks all conformal symmetries and supersymmetries.\footnote{The ``probe limit'' $g_s = 0$ only restores conformal symmetry and supersymmetry if all 7-branes are coincident \cite{Buchbinder}.  We will be mostly interested in the case of compact $\Sigma^2$, which involves 16 $D7$-branes and 4 $O7$-planes--ie. F-theory on $K3$.  Even in the orientifold limit, where $g_s$ is constant and may be set to zero everywhere, conformal symmetries are not restored, as the $O7$-planes can not be taken coincident, and the transverse space is compact, $\Sigma^2 = T^2/\mathbbm{Z}_2$.}

This supergravity solution was recently written down in \cite{Buchbinder}, where this system of intersecting branes was also studied.  That work considered the $D3/D7$ intersection as a surface operator in the field theory and studied its dual description.  We focus here on other aspects of the correspondence.

The next two sections contain the main results of the paper.  According to AdS/dCFT lore, defect operators localized on the brane intersection in the field theory should be dual to supergravity modes on the ``probe'' brane, coming from the massless excitations of open strings.  In our system the $D7$-branes are not probes; they produce their own backreaction on the geometry.  Nevertheless, we show in Section \ref{DefectSection} that when expanding the total action $S_{\textrm{bulk}} + S_{D7}$, where $S_{D7}$ is the sum of DBI and WZ actions for the $D7$-brane, around the background solution, the $D7$-brane action produces well-defined, finite couplings for open string modes.  The $D7$-branes lie on $AdS_3 \times S^5$ slices of the total geometry.  We then perform a Kaluza-Klein reduction on the five-sphere and obtain the complete bosonic and fermionic spectrum of $D7$-brane open string excitations on $AdS_3$.

These modes should be dual to operators on the field theory side that contain defect fields, involve a single color contraction, and fall into primary multiplets of the supersymmetry algebra.  If conformal supersymmetry was present, these would be short, chiral primary multiplets; since it is not, they are simply (long) multiplets of the ordinary supersymmetry algebra.  As such, there is no algebraic argument that protects the dimensions of these operators from receiving corrections as one goes to strong 't Hooft coupling.  Nonetheless, the mode-operator map that we propose combined with the $AdS_3/CFT_2$ mass-dimension relations imply that these operators are not renormalized.  It would be interesting to investigate this further from the field theory perspective.

In Section \ref{RSymmetrySection} we address the supergravity dual signature of the $D3$-brane zero-modes that localized to the intersection on the field theory side.  These modes are chiral and transform under the R-symmetry group of the field theory; in contrast, the modes of the 3-7 strings are singlets of the R-symmetry group.  Hence, there is a global anomaly of the R-symmetry current that localizes to the defect.  In the original analysis in \cite{Wittenads}, the anomaly of the R-symmetry current is reproduced in the supergravity description through a nonzero dual gauge field variation  that localizes to the boundary of AdS. We also find a term on the supergravity side whose gauge variation localizes to the boundary of $AdS_3$ and reproduces the proper anomaly; however its origin is somewhat of a surprise.

We make some concluding remarks in Section \ref{DiscussionSection}.

As pointed out in \cite{Buchbinder}, the proposed gauge/gravity duality for this system is somewhat novel.  There is a lack of decoupling of the 7-brane geometry on both sides of the correspondence.  Hence, the field theory is not in flat Minkowski space, and the five-dimensional gravity is not asymptotically $AdS_5$.  It is a correspondence between field theory in a supergravity background and string theory in a (different) supergravity background.

\section{Review of setup and field theory results}  \label{ReviewSection}

In this section we briefly review some key results of \cite{HR}.  The coordinate axes are taken such that the branes span the directions marked in Table \ref{table0}.
\TABLE{ \label{table0}
\begin{tabular}{l|c|c|c|c|c|c|c|c|c|c}
& 0 & 1 & 2 & 3 & 4 & 5 & 6 & 7 & 8 & 9 \\
\hline
$D7$, $O7$ & x & x & & & x & x & x & x & x & x \\
\hline
$D3$ & x & x & x & x & & & & & &  \\
\end{tabular}
\caption{Brane orientations}}

This intersection preserves at most a $Spin(1,1) \times Spin(2) \times Spin(6)$ subgroup of the ten-dimensional Lorentz group.  However, for the brane configurations we will consider, not all of the 7-branes are coincident, so the symmetry is reduced to $Spin(1,1) \times Spin(6)$.  The configuration is $1/4$ BPS, preserving eight of the IIB supercharges.  From the point of view of the $1+1$-dimensional intersection these are chiral fermions and we choose them to be right-handed.  Our conventions will be that four supercharges transform in the $(-1/2,-1/2, \bar{\mathbf{4}})$ of $Spin(1,1) \times Spin(2) \times Spin(6)$, and their conjugates in the $(-1/2,1/2,\mathbf{4})$.

\subsection{The 7-brane background}

For $g_s \neq 0$, the backreaction of the 7-branes on the geometry must be taken into account.  Further, since the $D3$-branes are extended in the directions transverse to the $D7$-branes, one is required to work with globally well-defined 7-brane supergravity solutions, and not simply local ones.  This leads one directly to the ``stringy cosmic string'' solutions of \cite{GSVY}, and their later interpretation in terms of $D7$-branes, $O7$-planes, and compactifications of F-theory to eight dimensions \cite{GGP,Vafa,Sen,Berg}.  We are interested in 7-brane solutions where the asymptotic value of the axidilaton is tunable and may be taken small--these are the solutions that can be embedded into a perturbative description of string theory.

The solutions involve a nontrivial metric and axidilaton.  Let $z = x^2 + i x^3$ parameterize the space transverse to the 7-branes.  Our conventions for the chirality of the preserved supercharges dictate that $\tau = C_0 + i e^{-\phi} \equiv \tau_1 + i \tau_2$ should be an anti-holomorphic function of $z$, or equivalently, $\bar{\tau} = \bar{\tau}(z)$.  The solution makes use of Klein's modular j-function, $j : F_0 \rightarrow \hat{\mathbbm{C}}$, a one-to-one and onto mapping of the fundamental domain of $PSL(2,\mathbbm{Z})$ to the Riemann sphere.  Specifically,
\begin{eqnarray} \bar{\tau}(z) & = & j^{-1}( \frac{P(z)}{Q(z)} ), \label{taueq} \end{eqnarray}
where $P(z),Q(z)$ are polynomials of the same degree, designed to make $\bar{\tau}$ regular at the cusps $\bar{\tau} = i,e^{2\pi i/3}$ of $F_0$.  We denote the degree of $Q$ by $N_f$, and the zeroes of $Q$ by $z_{i\infty}^{(n)}$.  Near these points $\bar{\tau} \rightarrow \frac{i}{2\pi} \ln{(z-z_{i\infty}^{(n)})}$; they correspond to the location of 7-branes.  In order for $P,Q$ to unwind the monodromy of $j^{-1}$ around $i,e^{2\pi i/3}$, we must have $N_f$ divisible by 6.  The solution for the ten-dimensional metric is
\begin{eqnarray}  ds^2 & = & \eta_{\mu\nu} dx^\mu dx^\nu + e^{a(z,\bar{z})} dz d\bar{z} + \delta_{IJ} dy^I dy^J, \quad \textrm{where} \nonumber \\
e^{a(z,\bar{z})} & = & \tau_2 \frac{ \eta^{2}(\bar{\tau}(z)) \bar{\eta}^2(\tau(\bar{z})) }{ \prod_{n = 1}^{N_f} (z - z_{i\infty}^{(n)} )^{1/{12}} ( \bar{z} - \bar{z}_{i\infty}^{(n)} )^{1/{12}} } \equiv \tau_2 g(z) \bar{g}(\bar{z}). \label{eaeq}  \end{eqnarray}
The factors of $(z-z_{i\infty}^{(n)})$ in the denominator cancel the zeroes of the Dedekind eta function at these points, so that the metric is everywhere nondegenerate.  Near the 7-branes it behaves as $\ln{|z-z_{i\infty}^{(n)}|}$, thanks to the factor of $\tau_2$.  As $z \rightarrow \infty$, the warp factor behaves as $e^{a} \rightarrow 1/|z|^{N_f/6}$.  Thus there is a deficit angle of $\delta = 2\pi \frac{N_f}{12}$.  The space is asymptotically conical, cylindrical, or is a two-sphere for $N_f = 6,12,24$ respectively.  These are the only values $N_f$ can take.  If we view $\tau$ as the modular parameter of a torus fibred over the base parameterized by $z,\bar{z}$, then these are solutions of F-theory.  In particular, the case where the base in a two-sphere corresponds to F-theory on $\mathbbm{R}^{1,7} \times K3$.

For definiteness, let us now discuss the $N_f = 24$ case.  Analogous statements to the following apply in the other two cases.  Though there are 24 7-branes, they can not all simultaneously be taken as perturbative $D7$-branes--ie. 7-branes on which $(p,q) = (1,0)$ strings end.  Some 7-branes are immersed in regions where the string coupling, $1/\tau_2$, is order one.  By applying an $SL(2,\mathbbm{Z})$ transformation one can make the string coupling small, but such a transformation acts nontrivially on the $(p,q)$ charges of the brane.  There is no globally perturbative description; rather, perturbative descriptions are patched together via $SL(2,\mathbbm{Z})$ transformations.  Though there is no globally perturbative description, one can make the regions of strong coupling arbitrarily small.  This is best understood by going to a special limit and considering small deformations around that limit.

The orientifold limit corresponds to taking certain combinations of $(p,q)$ 7-branes coincident such that the axidilaton is everywhere constant.  For the limit we are interested in, one takes the $z_{i\infty}^{(n)}$ to be equal in sets of six.  Each of the resulting four points is interpreted as the location of four $D7$-branes coincident with an $O7^{-}$-plane, such that the gauge group on their worldvolume is $SO(8)$.  The small deformations away from this limit consist of pulling the four $D7$-branes off of the $O7$-plane (in any of the four sets).  The $O7$-plane splits nonperturbatively into two $SL(2,\mathbbm{Z})$ transformed $(p,q)$ 7-branes, which are constrained to remain close to each other.  While the string coupling becomes order one in the vicinity of the $(p,q)$ 7-branes, the four ``satellite'' $D7$-branes act effectively as shields, keeping the string coupling small away from each system of six 7-branes.  By encircling each cloud of strong coupling while remaining in a perturbative description, we can measure its charge and monodromy to be that of an $O7^{-}$-plane, and therefore effectively treat it as such.  The $SL(2,\mathbbm{Z})$ monodromy around an $O7$-plane is $S^2 = - \mathbbm{1}$, where $T,S$ are the usual generators.  This has important consequences for the open strings on the $D3$-branes.

This system should be contrasted with studies of the AdS/CFT correspondence where flavor 7-branes are oriented in parallel to the $D3$-branes.  In the parallel case the 7-branes can be treated as probes \cite{KarchKatz,KMMW} or their backreaction can be taken into account \cite{GranaPolchinski,BLZY}.  The point is that in these situations one only requires a local description of the 7-brane background.  Therefore the number of 7-branes that one may consider is not restricted as it is here.

\subsection{Field theory symmetries and low energy effective action}

After expanding around the background above and taking the Maldacena $\alpha' \rightarrow 0$ low energy limit, the bulk and 7-brane fluctuations decouple from the system, leaving the effective low energy theory on the $D3$-branes and the intersection.

The field content on the $D3$-branes is that of $\mathcal{N}=4$ SYM, (though the low energy theory itself is not as we will explain).  We denote it as
\begin{eqnarray} (M^{ij}, \psi^i, A_m). \end{eqnarray}
Here the superscript $i,j = 1,\ldots,4$ is an index in the $\mathbf{4}$ of $SU(4)_R$ while subscript $i,j$ is an index in the $\bar{\mathbf{4}}$.  The three complex scalars have been packaged into an antisymmetric matrix $M^{ij} = -M^{ji}$ that additionally has a reality constraint $(M^{ij})^\dag = \frac{1}{2} \epsilon_{ijkl} M^{kl} \equiv M_{ij}$.  The fields are in the adjoint, but the issue of the gauge group is subtle.  While locally, away from the $D3/O7$ intersection it is $U(N_c)$, the presence of the $O7$-planes and their $SL(2,\mathbbm{Z})$ monodromy prevent the extension of this to a global symmetry group.  The action of $S^2$ on the open strings can be identified with worldsheet orientation reversal $\Omega$.  As this operation flips Chan-Paton labels, it is a generalized charge conjugation.  Thus one can identify the $D3/O7$ intersections as Alice strings \cite{Schwarz,BLeeP,BLoP}.  Only an $O(N_c)$ subgroup of the local symmetry group is globally well-defined.  We use $O(N_c)$ representations to classify field content, and the zero-modes to be mentioned below only take values in $O(N_c)$.

The low energy effective action on the $D3$-brane, which can be derived from the DBI and WZ actions \cite{Myers,MMS}, has the look of an $\mathcal{N} = 4$ action, but in the curved background $\mathbbm{R}^{1,1} \times \Sigma^2$ and with a spacetime varying Yang-Mills coupling and theta angle:
\begin{eqnarray}  S_{\textrm{3-3}} & = & S_{\textrm{3-3}}^{bos.} + S_{\textrm{3-3}}^{ferm.} \ , \\
S_{\textrm{3-3}}^{bos.} & = & - \frac{1}{2\pi} \int d^4 x \sqrt{-g} \textrm{tr} \displaystyle\biggl( \frac{1}{4} \tau_2 F_{mn} F^{mn} + \frac{1}{8} \tau_1 \epsilon^{mnpq} F_{mn} F_{pq} \nonumber \\
& & + \frac{1}{2} \mathcal{D}_m M_{ij} \mathcal{D}^m M^{ij} - \frac{1}{4 \tau_2} [ M_{ij} , M_{kl} ] [ M^{ij}, M^{kl} ] \displaystyle\biggr) \label{D3bosonic} \\
S_{\textrm{3-3}}^{ferm.} & = & \frac{i}{2\pi} \int d^4 x  \sqrt{-g} \textrm{tr} \displaystyle\biggl\{ \tau_2 \displaystyle\biggl( \bar{\psi}_i \gamma^m ( \mathcal{D}_m + \frac{i}{2} Q_m ) L \psi^i \displaystyle\biggr) \nonumber \\
& & +  \sqrt{\tau_2} \displaystyle\biggl( (L\psi)^i [ (L\psi)^j, M_{ij}] + (\bar{\psi} R)_i [ (\bar{\psi} R)_j, M^{ij}] \displaystyle\biggr) \displaystyle\biggr\}. \end{eqnarray}
We momentarily employ spacetime indices $m,n$ such that $x^m = (x^\mu, z, \bar{z})$.  The derivatives $\mathcal{D}_m$ are spacetime and gauge covariant.  $Q_m$ is a background $U(1)$ connection constructed from the axidilaton:
\begin{eqnarray} Q_m = \frac{\partial_m (\tau + \bar{\tau})}{2i (\tau - \bar{\tau})} \ . \label{U1connection} \end{eqnarray}
This is the same connection that appears in the IIB supergravity equations of motion.  It transforms as a gauge field under $SL(2,\mathbbm{Z})$ transformations.

The coupling of the fermions to the background $U(1)$ connection $Q_m$ has important consequences.  One can show via an index calculation that, in the $N_f = 24$ case, the fermions have chiral zero-modes that localize to the intersection.  There is a left-handed zero-mode that transforms in the antisymmetric tensor representation of $O(N_c)$ and a right-handed zero-mode that transforms in the symmetric tensor representation.  We denote these zero-modes $\xi_L,\xi_R$ respectively.  The rest of the $D3$-brane fields also have (nonchiral) zero-modes.  Their existence can be argued by going to the orientifold limit, where the brane configuration is T-dual to $N_c$ $D1$-strings in Type I.  However, they can also be explicitly constructed away from the orientifold limit \cite{HR}.

A standard exercise in string quantization shows that the massless modes of the 3-7 strings on the $D3/D7$ intersection consist of a single Weyl fermion in the bifundamental of $O(N_c) \times G_f$, where $G_f$ in the gauge group on the $D7$-brane worldvolume.  $G_f$ could be $SO(8)$, $U(4)$, or broken further, depending on the 7-brane locations.  This fermion, which we denote $q_L$, is a singlet under supersymmetry.  We take it to be left-handed, so that all eight preserved supercharges are right-handed.  In two dimensions our gamma matrix conventions are $\gamma_{(2)}^{0} = i \sigma^2, \gamma_{(2)}^{1} = \sigma^1$.  Symmetries dictate the form of the low energy action:
\begin{eqnarray} S_{\textrm{3-7}} & = & \frac{1}{2\pi} \int_{I} d^2 x q_{L}^\dag (i \partial_- - A_-) q_L \ , \label{37action} \end{eqnarray}
where $\partial_- = \frac{1}{2}(\partial_0 - \partial_1)$ etc.  Indices in $O(N_c) \times G_f$ are being suppressed.  There is a coupling to the $D7$-brane gauge field but in the $\alpha' \rightarrow 0$ limit this field becomes massive and $G_f$ becomes a global symmetry.  One can explicitly check that this action is invariant under the half of the $\mathcal{N}=4$ supersymmetries that are right-handed on the two-dimensional intersection.

The defect action given here is very simple in comparison to those studied in references \cite{DeWolfe,Erdmenger}, which included couplings of defect fields to normal derivatives of bulk scalars, for example.  The difference is that here the defect field content is purely fermionic and inert under supersymmetry.  On dimensional grounds the only other (marginal) defect-bulk coupling one could write down is of the form $q_{L}^\dag M^{ij} q_L$, but this respects neither supersymmetry nor R-symmetry.

In Table \ref{table1} we summarize the transformation properties of the chiral modes on the intersection and the supercharges under the various symmetry groups.
\TABLE{ \label{table1}
\begin{tabular}{l|l|l}
& $SO(1,1) \times \widehat{SO(2)} \times SO(6)$ & $O(N_c) \times G_f$ \\
\hline
$\xi_{L}^i$ & $(\frac{1}{2} , \frac{1}{2}, \mathbf{4})$ & $(\frac{ \mathbf{N_c} (\mathbf{N_c} - \mathbf{1} )}{ \mathbf{2}} ,\mathbf{1})$  \\
\hline
$\xi_{R}^i$ & $(-\frac{1}{2} , -\frac{1}{2}, \mathbf{4})$ & $(\frac{ \mathbf{N_c} (\mathbf{N_c} + \mathbf{1} )}{ \mathbf{2}} ,\mathbf{1})$  \\
\hline
$q_L$ & $(\frac{1}{2}, 0, \mathbf{1})$ & $(\mathbf{N}_c,\overline{fund.})$\\
\hline
$Q_{Ri}, Q_{R}^{\dag i}$ & $(-\frac{1}{2}, -\frac{1}{2}, \bar{\mathbf{4}}), (-\frac{1}{2}, \frac{1}{2}, \mathbf{4})$ & $(\mathbf{1},\mathbf{1})$
\end{tabular}
\caption{Transformation properties of chiral intersection modes and preserved supercharges.  We specify the charges of the fields under the action of the Abelian groups and the dimensions of their representations for non-Abelian groups.  The hat over $SO(2)$ is a reminder that it is not actually a symmetry of the background, though it is still useful to consider the charges of the fields under its action.}}

\section{Supergravity solution}  \label{GravitySection}

The effective field theory of the last section is an appropriate description of the system when the 't Hooft coupling is small, $\lambda \equiv g_s N_c << 1$.  Here, $g_s$ is related to the asymptotic value of the dilaton, $1/g_s = \lim_{z \rightarrow \infty} \tau_2$.  In the opposite regime, $\lambda >> 1$, one must account for the the backreaction of the $D3$-branes on the geometry.  Thus we require the full supergravity solution corresponding to the system of intersecting $D3$- and 7-branes.

This solution was recently written down in \cite{Buchbinder} and it is easily obtained from the ``harmonic function rule'' described in \cite{Tseytlin}.  We continue to use coordinates $(x^\mu, z, \bar{z})$, $\mu = 0,1$, for directions tangent to the $D3$-branes, and we now adopt spherical coordinates $(r,\theta^\alpha)$, $\alpha = 1,\ldots,5$, for directions transverse to the the $D3$-branes.  Then in Einstein frame the solution is given by
\begin{eqnarray} ds^2 & = & f(r)^{-1/2} [ \eta_{\mu\nu} dx^\mu dx^\nu + \lambda^2 e^{a(z,\bar{z})} dz d\bar{z} ] + f(r)^{1/2} [ dr^2 + r^2 d\Omega_{5}^2 ] \label{globalmetric} \\
\tau & = & \tau(\bar{z}) \\
G_{(5)} & = & (1 - \ast) 4 R^4 \epsilon_{S^5} \ , \label{RR5form} \end{eqnarray}
where
\begin{eqnarray} f(r) & =  & 1 + \frac{R^4}{r^4}  \end{eqnarray}
is the harmonic function from the $D3$-brane solution, $\epsilon_{S^5}$ is the volume form on the \emph{unit} 5-sphere, and $\tau, e^a$ are given by the GSVY solution \cite{GSVY}, equations \eqref{taueq},\eqref{eaeq} above.  One can easily generalize $f(r)$ to the multi-centered solutions describing separated $D3$-branes; we will restrict ourselves to the simplest case of coincident $D3$-branes here.  We use $G_{(p+2)}$ to denote R-R field strengths.  The radius\footnote{Note that since we have a spacetime varying dilaton, and we will be mostly considering the case where $e^a dz d\bar{z}$ describes a compact space, we do not bother pulling out the ``asymptotic'' value: $e^\Phi = g_s e^{\Phi - \Phi_0}$.  Hence there will be no factor of $g_s$ in front of the Hilbert action, and this why it does not appear in the formula for $R^4$.} $R$ is given by
\begin{eqnarray} R^4 & = & 4 \pi \alpha'^2 N_c \ , \label{Rvalue} \end{eqnarray}
and $\lambda$ is a free parameter that controls the size of the space, $\Sigma^2$, transverse to the 7-branes.  To be consistent with the conventions in the field theory analysis of the previous section, $\tau$ is again antiholomorpic.  Furthermore, since the ten-dimensional chirality of the IIB supercharges is taken to be right-handed, the five-form is taken to be anti-self-dual.

We are interested in doing AdS/CFT, so let us go to the near horizon of the $D3$-branes, letting $f(r) \rightarrow R^4/r^4$.  We have
\begin{eqnarray} ds^2 & \rightarrow & \frac{r^2}{R^2} \eta_{\mu\nu} dx^\mu dx^\nu + \frac{R^2}{r^2} dr^2 + \frac{r^2}{R^2} \lambda^2 e^{a} dz d\bar{z} + R^2 d\Omega_{5}^2 \nonumber \\
& = & ds_{AdS_3}^2 + \frac{r^2}{R^2} ds_{\Sigma^2}^2 + R^2 d\Omega_{5}^2 \ . \label{nearhorizon} \end{eqnarray}
Henceforth we will use the notation $x^m = (x^\mu, r)$ for the $AdS_3$ coordinates.  While the metric has an $AdS_3$ factor, it does not possess the full set of AdS isometries.  This is due to the warp factor in front of the $\Sigma^2$ term, which destroys the conformal symmetries.  The full bosonic isometry group is $ISO(1,1) \times SO(6)$.  Similarly, an analysis of the preserved supersymmetries of the background, given in Appendix \ref{PreservedSUSY}, shows that the eight supersymmetries of the global solution remain the only supersymmetries in the near-horizon region.  There is no doubling of the supersymmetry due to near-horizon conformal enhancement.

These statements match well with our expectations from the dual field theory.  The global bosonic symmetry of the effective theory on the $D3$-branes is $ISO(1,1) \times SO(6)$.  The preserved supersymmetries match, as we show in the appendix.  Finally, conformal symmetries and conformal supersymmetries are also broken in the field theory, due to the 7-brane geometry.

The first serious step in establishing the correspondence is to exhibit the mode-operator map.  We will restrict ourselves to the perturbative, $N_c \rightarrow \infty$, sector of the map--the relation between supergravity fluctuations about the background \eqref{nearhorizon} and single color-contraction operators falling into primary multiplets of the supersymmetry algebra.  This piece of the map can be further divided into two subsectors, defect and bulk.  In the next section we analyze the defect mode-operator map.

\section{Defect mode-operator map}  \label{DefectSection}

Applying the standard AdS/dCFT prescription to our system, field theory operators localized on the $D3/D7$ intersection, constructed using the 3-7 fermions $q_L$, should be dual to open string fluctuations on the $D7$-branes embedded in the near-horizon geometry \eqref{nearhorizon}.  Note that, while we have been careful to account for the backreaction of the 7-branes on the geometry, we are not \emph{replacing} the 7-branes by this geometry.

This is an important point and deserves further comment.  Indeed, one should never ``replace'' a brane by the geometry it produces.  When one analyzes the supergravity side of $AdS_{p+2}/CFT_{p+1}$, for instance, one is expanding in fluctuations around a background solution.  The solution is an extremum of $S_{II} + S_{Dp}^{static}$, where $S_{Dp}^{static}$ contains source terms for the metric, dilaton, and R-R form, $C_{p+1}$, on a static brane.  To be consistent then, one must expand $S_{II}$ in closed string fluctuations around the background geometry \emph{and} $S_{Dp}^{static}$ in open string fluctuations around the static brane.  However, for $p \leq 6$, the components of the Einstein metric in directions tangent to the brane involve a function, $f(r)$, of the radial distance from the brane, that vanishes at the source as $r \rightarrow 0$.  Thus the induced metric on the brane degenerates and open string fluctuations are removed from the spectrum.

The $D7$-brane is quite different and special in this respect.  The components of the metric tangent to the brane do not depend on the transverse coordinates and are completely well defined as one goes to the brane.  The following analysis shows that there are finite energy open string fluctuations on the $D7$-branes in the background \eqref{nearhorizon}.

\subsection{$D7$-brane fluctuations}

We begin with an expansion of the DBI and WZ actions in fluctuations, keeping lowest order in $\alpha'$.  The induced metric on the $D7$-brane worldvolume is that of $AdS_3 \times S^5$.  We then perform a Kaluza-Klein reduction on the five-sphere to obtain the mass spectrum on $AdS_3$.

For the bosonic $D7$-brane DBI and WZ actions, we follow the conventions of \cite{Myers}.  We choose the embedding coordinates to be the spacetime coordinates of $AdS_3 \times S^5$: $\xi^a =  (x^m, \theta^\alpha)$.  Labels $a,b,\ldots$ run over spacetime directions tangent to the brane.  Underlined indices $\underline{a},\underline{b}$, will denote corresponding tangent space directions.  Labels $i,j,\ldots$ run over the perpendicular directions: $x^i = (x^2,x^3)$ or $(z,\bar{z})$.  The bosonic $D7$-brane fields are the gauge field $A_a$ and the fluctuation scalars $\Phi^i$, which transform in the adjoint of the gauge group.  The gauge group could be $SO(8)$ in the orientifold limit, $U(4)$ away from the orientifold, or broken further.  The expansion of the DBI action in open and closed string fluctuations is straightforward.  We parameterize the fluctuations of the metric and dilaton around their background values as $G_{MN} = \tilde{G}_{MN} + h_{MN}$, $\phi = \tilde{\phi} + \delta \phi$; the background value of the NS-NS two-form $B_{MN}$ is of course zero.  We will denote the background induced metric on the brane as $g_{ab}$ .  In Einstein frame we find
\begin{eqnarray} S_{DBI}^{bos.} & = & - \mu_{D7} \int d^8\xi \sqrt{- g} e^{\tilde{\phi}} \textrm{tr} \displaystyle\biggl\{ 1 + \displaystyle\biggl(  \delta\phi + \frac{1}{2} h^{a}_{\phantom{a} a} + \frac{1}{2} (\delta\phi + \frac{1}{2} h^{a}_{\phantom{a} a} )^2  \nonumber \\
& & \qquad \qquad \qquad \qquad \qquad \qquad \qquad \quad - \frac{1}{4} h^{ab} h_{ab} + \frac{1}{4} e^{-\tilde{\phi}} B^{ab} B_{ab} \displaystyle\biggr) + \nonumber \\
& &+ (2\pi\alpha') \displaystyle\biggl( e^{-\tilde{\phi}/2} (h^{ai} + B^{ai}) \mathcal{D}_a \Phi_i + \frac{1}{2} e^{-\tilde{\phi}} B_{ab} F^{ab} \displaystyle\biggr) + \nonumber \\
& & + (2\pi\alpha')^2 \displaystyle\biggl( \frac{1}{2} e^{-\tilde{\phi}} \tilde{G}^{ij} \mathcal{D}_a \Phi_i \mathcal{D}^a \Phi_j + \frac{1}{4} e^{-\tilde{\phi}} F_{ab} F^{ab} \displaystyle\biggr) \displaystyle\biggr\} + O(\alpha'). \end{eqnarray}

There are three sets of terms.  The first set contains closed string modes only, the second set contains couplings between open and closed string modes, and the third set gives the open string couplings.  The terms displayed are those which, after canonically normalizing all fluctuations, survive the $\alpha' \rightarrow 0$ limit.  We will not consider the closed string couplings further, except to note that the one-point couplings must be cancelled by one-point couplings coming from the bulk supergravity action--we are expanding around a solution of the equations of motion.  As for the couplings between bulk and brane fields, we note that only the $U(1)$ brane fields survive the trace.  These couplings can have interesting consequences, but they will not affect the mass spectrum of open string modes.  We will only be interested in the third set of terms.

In order to determine the correct canonical normalization of the brane fields, we must first rescale the near-horizon metric appropriately.  The AdS/CFT correspondence requires that we zoom in on the $D3$-branes, $r \rightarrow 0$, while taking $\alpha' \rightarrow 0$ in such a way that $r/\alpha'$ remains fixed.  We define a new coordinate $v = r/R^2$, such that
\begin{eqnarray} \frac{R^2}{r^2} dr^2 + \frac{r^2}{R^2} \eta_{\mu\nu} dx^\mu dx^\nu &=& R^2 \displaystyle\biggl( \frac{dv^2}{v^2} + v^2 \eta_{\mu\nu} dx^\mu dx^\nu \displaystyle\biggr), \label{Mrescale} \end{eqnarray}
and we write $\tilde{G}_{MN} = R^2 \bar{G}_{MN}$.  After changing variables from $r$ to $v$, $\bar{g}_{ab}$ is the metric on $AdS_3 \times S^5$ with unit radius.  Now we replace $\tilde{G}$ in favor of $\bar{G}$.  One has $\sqrt{-g} F_{ab} F^{ab} = R^4 \sqrt{-\bar{g}} F_{ab} F^{ab}$ for instance, where on the right the indices are understood to be raised with the metric $\bar{g}^{ab}$.  Then the relevant terms from the DBI action are
\begin{eqnarray} S_{DBI}^{bos.} & \supset & - \frac{N_c}{8\pi^4} \int d^8 \xi \sqrt{-\bar{g}} \textrm{tr} \displaystyle\biggl\{ \frac{1}{4} F_{ab} F^{ab} + \frac{1}{2} \bar{G}^{ij} \mathcal{D}_a \Phi_i \mathcal{D}^a \Phi_j \displaystyle\biggr\},  \end{eqnarray}
where we have used \eqref{Rvalue} and $\mu_{D7} = (2\pi)^{-7} \alpha'^{-4}$.  The fact that all factors of $\alpha'$ cancel indicates that the $D7$-brane fluctuations are of finite energy in the Maldacena limit.

When the space transverse to the brane is curved, as is the case here, it is more natural to work with fluctuation scalars whose index is valued in the tangent space of the normal bundle.  Using the explicit form of the transverse metric and vielbeins, one finds
\begin{eqnarray} \bar{G}_{ij} \partial_a \Phi^i \partial^a \Phi^j & = & 2 \bar{G}_{z \bar{z}} \partial_a ( \bar{E}_{\underline{z}}^{\phantom{\underline{z}} z} \Phi^{\underline{z}} ) \partial^a ( \bar{E}_{\underline{\bar{z}}}^{\phantom{\underline{\bar{z}}} \bar{z}} \Phi^{\underline{\bar{z}}} ) \nonumber \\
& = & 2 \eta_{\underline{z} \underline{\bar{z}}} \displaystyle\biggl( \partial_{a} \Phi^{\underline{z}} \partial^{a} \Phi^{\underline{\bar{z}}} - v \partial_v (\Phi^{\underline{z}} \Phi^{\underline{\bar{z}}} ) + \Phi^{\underline{z}} \Phi^{\underline{\bar{z}}} \displaystyle\biggr) \ . \end{eqnarray}
When plugged back into the action, the second term can be integrated by parts.  Observe that $\partial_v ( v \sqrt{- \bar{g}} ) = \partial_v ( v^2 \sqrt{g_{S^5}} ) = 2 \sqrt{ - \bar{g}}$.  Hence this term will add to the third term.  Defining $\Phi \equiv \Phi^{\underline{z}}$, and so $\Phi^\ast = \Phi^{\underline{\bar{z}}}$, we obtain
\begin{eqnarray} S_{DBI}^{bos.} & \supset & - \frac{N_c}{8\pi^4} \int d^8 \xi \sqrt{-\bar{g}} \textrm{tr} \displaystyle\biggl\{ \frac{1}{4} f_{ab} f^{ab} +  \frac{1}{2} \partial_a \Phi  \partial^a \Phi^\ast + \frac{3}{2} |\Phi|^2 + int. \displaystyle\biggr\}, \end{eqnarray}
where $f_{ab}$ is the $O(A)$ part of $F_{ab}$.  In exchange for putting the kinetic term in standard form, we pick up a mass term for $\Phi$.  This will turn out be crucial for obtaining integer conformal dimensions from the $AdS_3/CFT_2$ mass-dimension relations.  Once we canonically normalize according to
\begin{eqnarray} (A_a, \Phi) & = & \frac{2^{3/2} \pi^2}{\sqrt{N_c}} (A_{a}' , \Phi'), \end{eqnarray}
the interaction terms are down by $1/\sqrt{N_c}$ or $1/N_c$.

The expansion of the bosonic WZ action to lowest order in $\alpha'$ also contains couplings of all three types: closed-closed, open-closed, and open-open string couplings.  We will not write out the full expression.  Fortunately, there is only one term of the last type.  It is given by
\begin{eqnarray} S_{WZ}^{bos.} & \supset & - \mu_{D7} \frac{(2\pi\alpha')^2}{2} \int \tilde{G}_5 \wedge \omega_3(A), \end{eqnarray}
where $d\omega_3(A) = \textrm{tr}(F^2)$, and we evaluate the five-form field strength on its background value \eqref{RR5form}, with all five legs taken along $S^5$.  Plugging this result in and replacing $A$ with the canonically normalized gauge field brings us to the simple result
\begin{eqnarray} S_{WZ}^{bos.} & \supset & - 2 \int \epsilon_{S^5} \wedge ( A' \wedge dA' + O( \frac{1}{\sqrt{N_c}} ) ). \end{eqnarray}

The fermionic $Dp$-brane actions in general bosonic supergravity backgrounds have been worked out very explicitly to quadratic order in fermions in \cite{MMS}.  They restricted attention to the Abelian case, but to quadratic order in fluctuations, which is all we will need, one can trivially generalize to the non-Abelian case by adding a trace.  Let us denote the fermion as $\Psi$.  It is a sixteen component spinor, but satisfies a Weyl constraint.  We take it to be left-handed: $L_{(8)} \Psi = \Psi$.  Then, after some work, one reduces the general formulae of \cite{MMS} to the following action, quadratic in fluctuations:
\begin{eqnarray} S^{ferm.} & = & \frac{N_c}{8\pi^4} \int d^8 \xi \sqrt{-\bar{g}} \textrm{tr}  \bar{\Psi} \displaystyle\biggl( i \gamma^a \nabla_a + \frac{1}{2} ( i \gamma^{\underline{01v}} - \gamma^{\underline{\theta^1 \cdots \theta^5}} ) \displaystyle\biggr) L_{(8)} \Psi \ . \label{FermionAction} \end{eqnarray}
The details of this calculation are presented in Appendix \ref{D7fermion}.  The factor out front is the same as we obtained for the bosonic action; this is required by supersymmetry, as we also discuss in the appendix.  $\nabla_a$ is the covariant derivative on $AdS_3 \times S^5$ with unit radius, and the $\gamma^a$ are $SO(1,7)$ gamma matrices.  The origin of the mass-like term is the coupling of the fermion to the background five-form field strength.

For the purpose of determining the spectrum of open string modes, the trace in all of these formulae may simply be replaced by a sum over an index valued in the adjoint of the 7-brane gauge group.  Hence, in the case of $SO(8)$ for instance, there are $8 \cdot 7/2$ copies of each field.  This index will always be suppressed in the following.

\subsection{K-K reduction on $S^5$}

We now expand the scalar, gauge field, and fermion in complete sets of appropriate spherical harmonics on $S^5$, and determine the resulting spectrum on $AdS_3$.  We begin with the scalar.  Dropping the primes from the normalization in the last section, the quadratic action is
\begin{eqnarray} S[\Phi] & = & - \int d^8 \xi \sqrt{-\bar{g}} \displaystyle\biggl( \frac{1}{2} \partial_a \Phi \partial^a \Phi^\ast + \frac{3}{2} |\Phi|^2 \displaystyle\biggr), \end{eqnarray}
leading to the equation of motion
\begin{eqnarray} \nabla_a \partial^a \Phi - 3 \Phi & = 0 \ . \end{eqnarray}

Now expand $\Phi$ in a complete set of scalar spherical harmonics on $S^5$:
\begin{eqnarray} \Phi & = & \sum_{I_1} \phi^{I_1}(x^m) Y^{I_1}(\theta^\alpha). \end{eqnarray}
We follow the notation and conventions of \cite{KRN}.  $I_1 = \{k,l_i,m_i\}$ with $k \geq 0$, and these harmonics satisfy $\Delta_{S^5} Y^{I_1} = -k (k+4) Y^{I_1}$.  The equation of motion for each mode on $AdS_3$ is then
\begin{eqnarray} \displaystyle\biggl( \nabla_m \partial^m - (k^2 + 4k + 3) \displaystyle\biggr) \phi^{I_1} & = & 0 \ , \qquad k \geq 0 \ . \end{eqnarray}
This is the equation for a scalar on $AdS_3$ with mass squared $m^2 = k^2 + 4k + 3$.

Next we consider the gauge field, with quadratic action
\begin{eqnarray} S[A] & = & - \int d^8 \xi \sqrt{-\bar{g}} \frac{1}{4} f_{ab} f^{ab} - 2 \int \epsilon_{S^5} \wedge A \wedge dA \ . \end{eqnarray}
Decompose the gauge field into $AdS_3$ and $S^5$ components, $A^a = (A^m, A^\alpha)$.  Then
\begin{eqnarray} f_{ab} f^{ab} & = & f_{mn} f^{mn} + f_{\alpha\beta} f^{\alpha\beta} +  2 f_{m\alpha} f^{m\alpha} \ , \qquad \textrm{and} \\
f_{m\alpha} f^{m\alpha} & = & \partial_m A_\alpha \partial^m A^\alpha + \partial_\alpha A_m \partial^\alpha A^m  - 2 \partial_\alpha A_m \partial^m A^\alpha \ . \end{eqnarray}
We make the gauge choice $\nabla_\alpha A^\alpha = 0$.  Then the last term in $f_{m\alpha} f^{m\alpha}$ vanishes after integration by parts.  Noting that the Chern-Simons term only involves $A_m$, the action separates as
\begin{eqnarray} S[A] & = & - \int d^8\xi \sqrt{-\bar{g}} \displaystyle\biggl( \frac{1}{4} f_{mn} f^{mn} + \frac{1}{2} \partial_{\alpha} A_m \partial^{\alpha} A^m \displaystyle\biggr) - 2 \int \epsilon_{S^5} \wedge A \wedge dA + \nonumber \\
& & - \int d^8\xi \sqrt{-\bar{g}} \displaystyle\biggl( \frac{1}{2} \partial_m A_\alpha \partial^m A^\alpha + \frac{1}{4} f_{\alpha\beta} f^{\alpha\beta} \displaystyle\biggr). \end{eqnarray}

Let us deal with the $S^5$ gauge field first.  Given our gauge choice, the equation of motion is
\begin{eqnarray} \nabla_m \partial^m A^\alpha + \textrm{Max}_{S^5} A^\alpha & = & 0 \ , \end{eqnarray}
where $\textrm{Max} A^\alpha = (\nabla_\gamma \nabla^\gamma \delta^{\alpha}_{\phantom{\alpha}\beta} - R^{\alpha}_{\phantom{\alpha}\beta} ) A^\beta$.  We expand $A^\alpha$ in vector spherical harmonics,
\begin{eqnarray} A_\alpha & = & \sum_{I_5} a^{I_5}(x) Y^{I_5}_{\alpha}(\theta), \end{eqnarray}
which have the properties $\nabla^\alpha Y_{\alpha}^{I_5} = 0$ and
\begin{eqnarray} \textrm{Max} Y_{\alpha}^{I_5} = - (k+1)(k+3) Y_{\alpha}^{I_5} \ , \qquad \textrm{for} \qquad k \geq 1 \ . \end{eqnarray}
Thus we get another tower of massive scalars on $AdS_3$ obeying the equations
\begin{eqnarray} \displaystyle\biggl( \nabla_m \partial^m - (k+1)(k+3) \displaystyle\biggr) a^{I_5} & = & 0 \ , \qquad k \geq 1 \ . \end{eqnarray}

Now consider the $AdS_3$ gauge field.  In this case we expand in spherical harmonics at the level of the action.  Let
\begin{eqnarray} A_m = \sum_{I_1} a^{I_1}_{m}(x) Y^{I_1}(\theta) \qquad \Rightarrow \qquad f_{mn} = \sum_{I_1} f_{mn}^{I_1} Y^{I_1} \ . \end{eqnarray}
Using orthonormality of the harmonics, we do the integral over the five-sphere and are left with
\begin{eqnarray} S & = & - \sum_{I_1} \int d^3x \sqrt{-g_{AdS}} \displaystyle\biggl( \frac{1}{4} f_{mn}^{I_1} f^{I_1 mn} + \frac{1}{2} m_{I_1}^2 a_{m}^{I_1} a^{I_1 m} + 2 \epsilon^{mnp} a_{m}^{I_1} \partial_n a_{p}^{I_1} \displaystyle\biggr), \end{eqnarray}
where $m_{I_1}^2 = k(k+4)$, for $k \geq 0$, and $\epsilon^{01v} = (-g_{AdS})^{-1/2}$.  This is an example of a 3-dimensional Proca-Chern-Simons theory \cite{PisarskiRao,DeserTekin,MincesRivelles}.  The second term is a standard Proca mass term for the gauge field and the last term is a ``topological mass'' term.  The propagator for the photon of this theory has two different poles, both physical.  If we denote the coefficient in front of the Chern-Simons term by $\kappa/2$, so that in our case $\kappa = 4$, then these poles are located at
\begin{eqnarray} m_{\pm} & = & \frac{1}{2} (\sqrt{ \kappa^2 + 4 m_{I_1}^2} \pm | \kappa | ). \end{eqnarray}
Thus we find
\begin{eqnarray} m_{\pm} = \frac{1}{2} ( \sqrt{16 + 4 k (k+4) } \pm 4) = \left\{ \begin{array}{c} k + 4 \\ k \end{array}\right. , \qquad k \geq 0 \ . \end{eqnarray}
At each of these poles there is only one physical degree of freedom; the polarization vector has one free component.  In terms of the asymptotic Minkowski boundary, the excitations are left- or right-moving.  Let $a^{\pm}_{\mu}$ denote the boundary components of the modes corresponding to $m_{\pm}$.  Then, converting the results of \cite{MincesRivelles} to Minkowski signature $AdS_3$, one finds $a^{\pm}_{\mu} = \mp \varepsilon^{\mu\nu r} a^{\pm}_{\nu}$, where $\varepsilon^{01r} = 1$.  It follows that $a^{+}_{\mu}$ is right-moving, while $a^{-}_{\mu}$ is left-moving.

Finally, let us consider the $D7$-brane fermion action, given by
\begin{eqnarray} S[\Psi] & = & \int d^8 \xi \sqrt{-\bar{g}} \textrm{tr}  \bar{\Psi} \displaystyle\biggl( i \gamma^a \nabla_a + \frac{1}{2} ( i \gamma^{\underline{01v}} - \gamma^{\underline{\theta^1 \cdots \theta^5}} ) \displaystyle\biggr) L_{(8)} \Psi \ , \label{D7fa} \end{eqnarray}
after canonically normalizing.  We decompose the $SO(1,7)$ gamma matrices according to $SO(1,7) \rightarrow SO(1,2) \times SO(5)$ via the following:
\begin{eqnarray} \gamma^m & = & \sigma^1 \otimes \tau^m \otimes \mathbbm{1}_4 \ , \qquad m = 0,1,v \ , \\
\gamma^{\theta^\alpha} & = & -\sigma^2 \otimes \mathbbm{1}_2 \otimes \rho^{\alpha} \ , \qquad \alpha = 1, \ldots, 5 \ , \end{eqnarray}
where
\begin{eqnarray} \{ \tau^{\underline{m}} , \tau^{\underline{n}} \} = 2 \eta^{\underline{mn}} \ , \qquad \{ \rho^{\underline{\alpha}} , \rho^{\underline{\beta}} \} = \delta^{\underline{\alpha\beta}} \ . \end{eqnarray}
We will take $\tau^{\underline{0}} = i \sigma^2$, $\tau^{\underline{1}} = \sigma^1$, and $\tau^{\underline{v}} = \sigma^3$.  Further, we choose $\rho^{\underline{5}} = - \rho^{\underline{1}} \cdots \rho^{\underline{4}}$, so that $\prod \rho^{\underline{\alpha}} = - \mathbbm{1}$.  With these conventions the $SO(1,7)$ ``$\gamma^5$'' is given by
\begin{eqnarray} \bar{\gamma} = i \prod \gamma^{\underline{a}} = i \sigma^1 \sigma^2 \otimes \mathbbm{1}_2 \otimes \mathbbm{1}_4 = \left(\begin{array}{c c} - \mathbbm{1}_8 & 0 \\ 0 & \mathbbm{1}_8 \end{array}\right). \end{eqnarray}
Therefore we write
\begin{eqnarray} L_{(8)} \Psi & = & \left( \begin{array}{c} 0 \\ \psi \end{array}\right), \end{eqnarray}
where $\psi$ is an eight-component complex spinor.  In this basis we also find
\begin{eqnarray}  \frac{1}{2} ( i \gamma^{\underline{01v}} - \gamma^{\underline{\theta^1 \cdots \theta^5}} ) = \frac{1}{2}( i \sigma^1 - \sigma^2) \otimes \mathbbm{1}_8 = \left( \begin{array}{c c} 0 & i \mathbbm{1}_8 \\ 0 & 0 \end{array}\right). \end{eqnarray}
Lastly,
\begin{eqnarray} i \gamma^a \nabla_a & = & i \sigma^1 \otimes \tau^m \nabla_m \otimes \mathbbm{1}_4 - i \sigma^2 \otimes \mathbbm{1}_2 \otimes \rho^\alpha \nabla_\alpha \nonumber \\
& = & \left( \begin{array}{c c} 0 & i \slashed{\nabla}_x \otimes \mathbbm{1}_4 - \mathbbm{1}_2 \otimes \slashed{\nabla}_\theta \\ i \slashed{\nabla}_x \otimes \mathbbm{1}_4 + \mathbbm{1}_2 \otimes \slashed{\nabla}_\theta & 0 \end{array}\right), \end{eqnarray}
where we are using the shorthand $\slashed{\nabla}_x = \tau^m \nabla_m$ and $\slashed{\nabla}_{\theta} = \rho^\alpha \nabla_\alpha$.  Thus the equation of motion that follows from the action \eqref{D7fa} is
\begin{eqnarray} ( i \slashed{\nabla}_x \otimes \mathbbm{1}_4 - \mathbbm{1}_2 \otimes \slashed{\nabla}_\theta + i \mathbbm{1}_8 ) \psi & = & 0 \ . \end{eqnarray}

We expand $\psi$ in a complete set of spinor spherical harmonics,
\begin{eqnarray} \psi & = & \sum_{I_L} \lambda^{I_L}(x) \otimes \Theta^{I_L}(\theta), \end{eqnarray}
with $I_L = (\pm, k)$, $k \geq 0$, and satisying
\begin{eqnarray} \slashed{\nabla}_\theta \Theta^{I_L} = m_{I_L} \Theta^{I_L} \ , \qquad \textrm{with} \qquad m_{\pm,k} = \mp i (k + \frac{5}{2}). \end{eqnarray}
Note that $\Theta^{\pm,k}$ can be expressed in terms of the $k^{th}$ scalar spherical harmonic and the Killing spinor $\eta^{\pm}$ on $S^5$.  The Killing spinors satisfy $\nabla_\alpha \eta^{\pm} \pm \frac{i}{2} \rho_{\alpha} \eta^{\pm} = 0$, and are related by conjugation.  Plugging this expansion into the equation of motion, we obtain the following set of Dirac equations on $AdS_3$:
\begin{eqnarray}  ( \slashed{\nabla}_x + \tilde{m}_{I_L} ) \lambda^{I_L} = 0 \ , \qquad \textrm{with} \qquad \tilde{m}_{\pm,k} = [ \pm (k + \frac{5}{2}) + 1], \quad k \geq 0 \ .  \end{eqnarray}
The sign of the mass in this equation has important consequences for the asymptotic behavior of the spinor \cite{Henningson,LeighRozali}.  One can decompose the spinor according to $\lambda = \lambda_L + \lambda_R$, where $\lambda_{L,R}$ are eigenvectors of $\tau^{\underline{v}}$: $\tau^{\underline{v}} \lambda_{L} = \lambda_{L}$ and $\tau^{\underline{v}} \lambda_{R} = - \lambda_{R}$.  On the two-dimensional boundary $\tau^{\underline{v}}$ has the interpretation of the chirality operator.  In order for $\lambda$ to scale appropriately to some boundary data $\lambda_{\infty}$, in accordance with the AdS/CFT prescription, one of the $\lambda_{\infty L,R}$ must be set to zero.  The sign of the mass determines which one.  For $m >0$ the boundary data is right-handed, while for $m < 0$ it is left-handed.  In summary, we have the following towers of fermionic modes:
\begin{eqnarray} \begin{array}{l l}\lambda^{(+,k)}_{R} \ , & \tilde{m}_{+,k} = \frac{7}{2}, \frac{9}{2}, \ldots \\
\lambda^{(-,k)}_{L} \ , & \tilde{m}_{-,k} = -\frac{3}{2}, - \frac{5}{2}, \ldots  \end{array} . \end{eqnarray}
As we go to the boundary, each of these becomes a complex Weyl fermion of the indicated chirality.
\TABLE{ \label{table2}
\begin{tabular}{l|l|l}
Mode & mass (squared) & $SU(4)$ Dynkin labels \\
\hline
$a_{-}^{(k)}$, $k \geq 0$ & $k^2$ & $(0,k,0)$ \\
\hline
$\lambda^{(-,k)}_{L}$, $k \geq 0$ & $-(k + 3/2)$ & $(0,k,1)$ \\
\hline
$\phi^{(k)}$, $k \geq 0$ & $k^2 + 4k + 3$ & $(0,k,0)$ \\
\hline
$a^{(k)}$, $k \geq 1$ & $k^2 + 4k + 3$ & $(1,k-1,1)$ \\
\hline
$\lambda^{(+,k)}_R$, $k \geq 0$ & $(k + 7/2)$ & $(1,k,0)$ \\
\hline
$a_{+}^{(k)}$, $k \geq 0$ & $(k+4)^2$ & $(0,k,0)$
\end{tabular}
\caption{The spectrum of $D7$-brane open string modes on $AdS_3$.  The $\phi$ are modes of the complex transverse fluctuation scalar, the $a_{\pm}$ are modes of the $AdS_3$ gauge field, the $a$ come from the $S^5$ gauge field, and the $\lambda$ from the $D7$-brane fermion.  We give the mass-squared in the case of bosons and the mass in the case of fermions.  Note that there are $dim{G_f}$ copies of each field on a given set of coincident 7-branes, where $G_f$ is the corresponding gauge group.}}

This completes the K-K reduction of the $D7$-brane open string modes.  The results are summarized in Table \ref{table2}.  Additionally we give the $SU(4)$ representation in which each mode transforms.  These can be obtained by transcribing the appropriate $SO(6)$ representations which, in turn, are discussed in \cite{Rubin}.  For example, the $k^{th}$ scalar spherical harmonic transforms in the symmetric product of $k$ $SO(6)$ vectors.  A vector of $SO(6)$ corresponds to an antisymmetric product of two fundamentals in $SU(4)$; the Dynkin label would be $(0,1,0)$.  Thus the $k^{th}$ scalar spherical harmonic transforms in the $(0,k,0)$ of $SU(4)$.  The $k=1$ vector spherical harmonic transforms in the antisymmetric product of two $SO(6)$ vectors.  This corresponds to $(1,0,1)$ in $SU(4)$.  The $k^{th}$ vector spherical harmonic transforms in the $(1,k-1,1)$.  The Killing spinors $\eta^{\pm}$ transform in the fundamental and anti-fundamental of $SU(4)$.  Hence the $\Theta^{\pm,k}$ transform in the $(1,k,0)$ and $(0,k,1)$.

\subsection{The defect operators and the matching}

The open string modes discussed above should be dual to defect operators localized on the $D3/D7$ intersection in the field theory.  These operators must include the defect 3-7 fermion, $q_L$, and may include $D3$-brane fields as well, evaluated on the intersection.

In the usual correspondence, gauge invariance dictates that operators must be products of traces of adjoint valued fields.  The trace is the only $SU(N_c)$ invariant when one has only adjoint valued fields to work with.  The $q_L$, on the other hand, are in the fundamental of $O(N_c)$.  This gives us a second type of invariant to work with: $q_{L}^\dag \varphi^a T^a q_L$.  Here $\varphi^a$ is any adjoint valued field (or product of fields), and $T^a$ are the generators of $O(N_c)$ in the fundamental representation.  In the usual correspondence, the 't Hooft large $N_c$ limit of correlation functions tells one that multiple trace operators are suppressed by powers of $1/N_c$ from single trace operators.  Similarly, the 't Hooft analysis of the case with fundamental fields allows one to conclude that multiple q-contractions are suppressed by powers of $1/\sqrt{N_c}$ from single ones.

In the case of coincident 7-branes, the corresponding gauge group, $G_f$, becomes a global symmetry group of the field theory in the limit we are considering.  $q_L$ is in the fundamental of $G_f$.  Thus, the type of operators discussed above can also carry an index in the adjoint of $G_f$, $\mathcal{O}^A = q_{L}^\dag T^{aA} \varphi^a q_L$, where now $T^{aA}$ are generators of $O(N_c) \times G_f$.  The same index is carried by the open string supergravity modes on the coincident 7-branes.  They match in the obvious way and will henceforth be suppressed.

Supergravity modes should correspond to primaries of the supersymmetry algebra.  The multiplets are filled out by acting on a lowest weight operator with the preserved supercharges.  In direct analogy with the standard correspondence, we claim the the complete set of defect (single q-contraction) lowest weight operators is
\begin{eqnarray} \mathcal{O}^{(n)} & \equiv & q_{L}^\dag M^{ \{ I_1} \cdots M^{ I_n \} } q_L \ , \qquad n = 0,1,2,\ldots \ . \end{eqnarray}
Here $I$ is an index in the $\mathbf{6}$ of $SU(4)_R$ and $M^I$ are the $D3$-brane scalars, related to the $M^{ij}$ appearing in the effective action \eqref{D3bosonic} through $SU(4)$ Clebsch-Gordan coefficients for $\mathbf{4} \times \mathbf{4} \rightarrow \mathbf{6}$.  The curly brackets indicate the totally traceless symmetric product.  If we had superconformal symmetry, these would be (defect) chiral primary operators, in analogy with the operators considered in \cite{DeWolfe}.  Under the preserved subalgebra of ordinary supersymmetry, the multiplets built on $\mathcal{O}^{(n)}$ by acting with supercharges simply become long multiplets.

Now we need to determine all of the combinations of preserved supercharges that can act on $\mathcal{O}^{(n)}$ without annihilating it or producing a descendant.  The $q_L$ are singlets under the preserved supersymmetry, so the supercharges act only on the $D3$-brane fields.  Thus the situation is very similar to the standard $AdS_5$--$\mathcal{N} = 4$ SYM correspondence, except that we only have half of the sixteen $\mathcal{N}=4$ supercharges to work with.  As was pointed out in \cite{HR}, the supercharges that leave the 3-7 string action \eqref{37action} invariant are $Q_{2j},\bar{Q}_{\dot{2}}^{j}$, in the standard $d=4$ Wess and Bagger notation.  Their anti-commutator is $\{ Q_{2j} , \bar{Q}_{\dot{2}}^k \} = 2 P_- \delta_{j}^{k}$, the right-moving momentum along the intersection.  Since the preserved $Q$ all have the same value of the $SO(1,3)$ Weyl index, from the $\mathcal{N}=4$ point of view, one can only consider products of $Q$'s symmetrized on this index.  This is the principle restriction.  In the following we simply denote $Q_{2j} \equiv Q_j$ and $\bar{Q}_{\dot{2}}^j \equiv Q^{\dag j}$.

The operator $\mathcal{O}^{(n)}$ clearly transforms in the $(0,n,0)$ of $SU(4)$.  Acting with one supercharge we can have $Q_{j} \mathcal{O}^{(n)}$ or $Q^{\dag j} \mathcal{O}^{(n)}$ transforming in the $(1,n-1,0)$ and $(0,n-1,1)$ respectively.  Acting with two supercharges, there are three possibilities.  $Q_i Q_j \mathcal{O}^{(n)}$ gives a complex field in the $(0,n-1,0)$ and $Q^{\dag i} Q^{\dag j} \mathcal{O}^{(n)}$ its conjugate.  On the other hand $Q_{i} Q^{\dag j} \mathcal{O}^{(n)}$ gives a real field in the $(1,n-2,1)$.  Acting with three supercharges gives two possibilities.  $Q_i Q^{\dag j} Q^{\dag k} \mathcal{O}^{(n)}$ is in the $(1,n-2,0)$, and $Q^{\dag i} Q_j Q_k \mathcal{O}^{(n)}$ is in the $(0,n-2,1)$.  A product of three $Q$'s or their conjugates must annihilate the operator.  Finally, with four supercharges there is one possibility: $Q_i Q_j Q^{\dag k} Q^{\dag l} \mathcal{O}^{(n)}$ is in the $(0,n-2,0)$ of $SU(4)_R$.  Note that the $n=0,1$ multiplets are short.
\TABLE{ \label{table3}
\begin{tabular}{l|l|l}
Operator & $(\Delta_L, \Delta_R)$ & $SU(4)$ Dynkin labels \\
\hline
$\mathcal{O}^{(n)}$  & $(1+n/2,n/2)$ & $(0,n,0)$ \\
\hline
\multicolumn{3}{l}{$n \geq 1$} \\
\hline
$Q_i \mathcal{O}^{(n)}$  & $(1+n/2,1/2 + n/2)$ & $(1,n-1,0)$ \\
\hline
$Q^{\dag i} \mathcal{O}^{(n)}$  & $(1+n/2,1/2 + n/2)$ & $(0,n-1,1)$ \\
\hline
$Q_i Q_j \mathcal{O}^{(n)}$, $Q^{\dag i} Q^{\dag j} \mathcal{O}^{(n)}$  & $(1 + n/2,1 + n/2)$ & $(0,n-1,0)$ \\
\hline
\multicolumn{3}{l}{$n \geq 2$} \\
\hline
$Q_i Q^{\dag j} \mathcal{O}^{(n)}$, & $(1 + n/2, 1 + n/2)$ & $(1,n-2,1)$ \\
\hline
$Q_i Q^{\dag j} Q^{\dag k} \mathcal{O}^{(n)}$, & $(1 + n/2, 3/2 + n/2)$ & $(1,n-2,0)$ \\
\hline
$Q^{\dag i} Q_j Q_k \mathcal{O}^{(n)}$, & $(1 + n/2, 3/2 + n/2)$ & $(0,n-2,1)$ \\
\hline
$Q_i Q_j Q^{\dag k} Q^{\dag l} \mathcal{O}^{(n)}$, & $(1 + n/2, 2 + n/2)$ & $(0,n-2,0)$
\end{tabular}
\caption{Defect primary multiplets.}}

In Table \ref{table3} we summarize these results.  We also give the (classical) left and right weights of the operators, from the $1+1$-dimensional point of view.  Note that $q_L,q_{L}^\dag$ each have weight $(\Delta_L, \Delta_R) = (1/2,0)$, while each supercharge comes with weight $(0,1/2)$.  The $1+3$-dimensional bulk scalars are nonchiral and have total weight $1$; thus they contribute $(1/2,1/2)$ each.  Since these are long multiplets of the preserved supersymmetry algebra, there is no algebraic argument that protects the classical operator dimensions $\Delta = \Delta_L + \Delta_R$ from being corrected as we go to strong 't Hooft coupling.  Nevertheless, the following correspondence suggests that the dimensions are not renormalized.  Let us first present the map and then return to this point.

After staring at Tables \ref{table2},\ref{table3} for a little while one can conjecture the mode-operator map presented in Table \ref{table4}.  The $SU(4)$ quantum numbers match.  Furthermore the $SO(1,1)$ spin eigenvalues, given by $\Delta_L - \Delta_R$ in the operator case, are consistent.  Recall that $a_-$ corresponds to a left-moving mode on the Minkowski boundary of $AdS_3$, while $a_+$ corresponds to a right-moving mode.  Also, $\lambda_{L}^{(-,k)}$ is a left-handed complex Weyl spinor on the boundary, as is its conjugate, $\bar{\lambda}_{L}^{(+,k)}$.  (Recall here that the $\pm$ indicates the $SU(4)$ representation--fundamental or antifundamental).  Similarly, $\lambda_R, \bar{\lambda}_R$ are right-handed on the boundary.
\TABLE{ \label{table4}
\begin{tabular}{|l|l|l|}
\hline
& Operator & SUGRA mode  \\
\hline
$n = 0$ & $\mathcal{O}^{(0)} = q_{L}^\dag q_L$ & $a_{-}^{(k =0)}$  \\
\hline
$n=1$ & $\mathcal{O}^{(1)} = q_{L}^\dag M^{I} q_L$ & $a_{-}^{(k=1)}$  \\
\cline{2-3}
& $Q_i \mathcal{O}^{(1)}$, $Q^{\dag i} \mathcal{O}^{(1)}$ & $\bar{\lambda}_{L}^{(+,k=0)}$, $\lambda_{L}^{(-,k=0)}$ \\
\cline{2-3}
& $Q_i Q_j \mathcal{O}^{(1)}, Q^{\dag i} Q^{\dag j} \mathcal{O}^{(1)}$ & $\phi^{(k=0)}, \phi^{\ast (k=0)}$  \\
\hline
$n \geq 2$ & $\mathcal{O}^{(n)}$ & $a_{-}^{(k=n)}$  \\
\cline{2-3}
& $Q_i \mathcal{O}^{(n)}$, $Q^{\dag i} \mathcal{O}^{(n)}$ & $\bar{\lambda}_{L}^{(+,k=n-1)}$, $\lambda_{L}^{(-,k=n-1)}$ \\
\cline{2-3}
& $Q_i Q_j \mathcal{O}^{(n)}, Q^{\dag i} Q^{\dag j} \mathcal{O}^{(1)}$ & $\phi^{(k=n-1)}, \phi^{\ast (k=n-1)}$  \\
\cline{2-3}
& $Q_i Q^{\dag j} \mathcal{O}^{(n)}$ & $a^{(k=n-1)}$  \\
\cline{2-3}
& $Q_i Q^{\dag j} Q^{\dag k} \mathcal{O}^{(n)}$, $Q^{\dag i} Q_j Q_k \mathcal{O}^{(n)}$ & $\lambda_{R}^{(+,k=n-2)}$, $\bar{\lambda}_{R}^{(-, k=n-2)}$ \\
\cline{2-3}
& $Q_i Q_j Q^{\dag k} Q^{\dag l} \mathcal{O}^{(n)}$ & $a_{+}^{(k=n-2)}$ \\
\hline
\end{tabular}
\caption{The defect mode-operator map.}}

The most important ``check'' is the relation between the masses of the modes and the dimensions of the operators, $\Delta = \Delta_L + \Delta_R$.  The mass-dimension relations for $AdS_{d+1}$ scalars, spinors, and vectors are the following:
\begin{eqnarray} \textrm{scalars:} \qquad \Delta & = & \frac{1}{2} (d + \sqrt{d^2 + 4 m^2} ), \\
\textrm{spinors:} \qquad \Delta & = & \frac{1}{2} (d + 2 |m| ), \\
\textrm{vectors:} \qquad \Delta & = & \frac{1}{2} (d + \sqrt{ (d-2)^2 + 4 m^2} ). \end{eqnarray}
Setting $d=2$ and plugging the masses of Table \ref{table2} into the appropriate formulae, we indeed find the dimensions of the proposed dual operators in all cases.  This is no check for us, but rather a prediction of the correspondence.  The operator dimensions in these formulae are to be evaluated at strong 't Hooft coupling, so this result implies that the classical dimensions do not receive any corrections.  This is surprising because, as we have emphasized, the system does not possess superconformal symmetry and the defect operators are in ordinary long multiplets of the supersymmetry algebra.

As discussed in \cite{Buchbinder}, there is a probe ``limit'' where $g_s = 0$ and all 7-branes are taken coincident.  Then superconformal symmetry is present, with supergroup $SU(1,1|4)$.  In such a system, the defect operators discussed above are in short multiplets of the symmetry algebra and their dimensions are protected.  However, this configuration is not in the moduli space of F-theory solutions that we are considering.  Even in the orientifold limit of the system we study, the separation of the four $D7/O7$-planes and the compactness of the transverse space break conformal symmetry.  And certainly in the more general curved case, there is no clear reason to expect the operator dimensions to be protected.  It would be quite interesting to investigate this issue further from the field theory side, but we leave this for future work.

\section{Localized R-symmetry anomaly from the chiral $D3$-brane zero-modes}   \label{RSymmetrySection}

In the previous section we presented evidence from the defect sector for the proposed gauge/gravity duality.  In the analysis of other AdS/dCFT systems, where the defect branes are treated as probes, the bulk sector of the correspondence is simply equivalent to the standard AdS/CFT correspondence.  This is not the case here.  After reduction on the five-sphere, the bulk five-dimensional geometry is not $AdS_5$, or even asymptotically $AdS_5$.  Correspondingly, the boundary of that space, on which the dual field theory lives, is not simply four-dimensional Minkowski space.

One of the most interesting consequences of having the field theory defined on the curved background $\mathbbm{R}^{1,1} \times \Sigma^2$, is the presence of chiral zero-modes in the case of compact $\Sigma^2$.  These are zero-modes of the $D3$-brane fermion that localize to the intersection of the $D3$-branes with the 7-branes.  They are the $\xi_{L}^i$ and $\xi_{R}^i$ of Table \ref{table1}.  In contrast to the 3-7 fermion $q_L$, these modes transform under the $SO(6) \simeq SU(4)$ R-symmetry group.  The $i$ is an index in the $\mathbf{4}$.  Since there is an unequal number of left- and right-handed modes--the $\xi_{L}^i$ transform in the antisymmetric tensor of $O(N_c)$ while the $\xi_{R}^i$ transform in the symmetric tensor--there will be a global anomaly in the R-symmetry current \emph{that localizes} to the $D3/D7$-$O7$ intersection.  From the perspective of the two-dimensional field theory, these zero-modes will produce an $SU(4)$ current algebra at level $N_c$.  If the proposed duality is correct, we must be able to see the manifestation of this anomaly on the supergravity side.

In the standard $\mathcal{N}=4$ SYM there is also an R-symmetry anomaly.  The $AdS_5$ gauge field of $\mathcal{N}=8$ gauged supergravity is dual to the R-symmetry current.  The global anomaly in the field theory is reproduced in the supergravity by the nonzero gauge variation of a five-dimensional Chern-Simons term that localizes to the boundary of $AdS_5$ \cite{Wittenads}.  The $D3/D7$-$O7$ intersection should be identified with the asymptotic boundary of the 7-branes in the near-horizon geometry.  What we need is a three-dimensional Chern-Simons term on the $AdS_3$ of the 7-brane worldvolume.  Specifically, we would like to generate the required term according the mechanism discussed in \cite{KrausLarsen,HansenKraus}, where the holographic interpretation of anomalies in the $AdS_3/CFT_2$ context was well understood.

Let us now briefly review the essentials of this mechanism for the case of $AdS_3 \times S^5$.  From the higher dimensional perspective, the $SO(6)$ supergravity gauge field comes from a nontrivial connection on the five-sphere.  Let $y^{I}(\theta)$ be such that
\begin{eqnarray} \delta_{IJ} y^{I} y^{J} = 1 \ , \qquad \delta_{I J} dy^I dy^J = d\Omega_{5}^2 \ . \end{eqnarray}
Introduce one-forms on the $AdS_3$ base $\mathcal{A}^{IJ} = \mathcal{A}_{m}^{\phantom{m}IJ} dx^m$, $\mathcal{A}^{IJ} = - \mathcal{A}^{JI}$, and consider the coordinate transformation of the metric
\begin{eqnarray} ds_{AdS_3}^2 + d\Omega_{5}^2 \qquad \rightarrow \qquad ds_{AdS_3}^2 + (dy^I - \mathcal{A}^{IJ} y^J )^2 \ . \label{gaugemetric} \end{eqnarray}
One can show that $\mathcal{A}^{IJ}$ plays the role of a spin connection.  Then, with this metric, the gravitational Chern-Simons term on $AdS_3$ produces the gauge Chern-Simons term in addition to the usual piece.  Specifically, a term of the form
\begin{eqnarray} S_{c.s.}  =  \frac{k}{4\pi} \int_{AdS_3} \omega_3( \mathcal{A} )  , \qquad \textrm{with} \qquad \omega_3(\mathcal{A}) = \textrm{tr}_{\mathbf{4}} (\mathcal{A} d \mathcal{A} + \frac{2}{3} \mathcal{A}^3), \label{whatwewant}  \end{eqnarray}
will produce an anomaly matching that of the chiral zero-modes provided that we identify $k$ with the level of the current algebra, $N_c$.  Thus our task is to produce the gravitational Chern-Simons term on $AdS_3$, with coefficient $N_c/4\pi$, from the supergravity action for the system.

There are two different places we might find such a term.  The $AdS_5$ Chern-Simons term comes from the dimensional reduction of IIB supergravity around the $AdS_5 \times S^5$ solution.  It could be that in order to obtain the required term we must expand IIB supergravity in fluctuations around the solution of Section \ref{GravitySection}, and then dimensionally reduce on both the $S^5$ and compact $\Sigma^2$.  This could perhaps be done.  Fortunately there is another place where we can find such a term, and the analysis is far simpler.  There are bulk Chern-Simons couplings in the 7-brane WZ action.

Consider the WZ action for \emph{one set} of $4D7+O7$'s.  There is a term in the action, surviving the $\alpha' \rightarrow 0$ low energy limit, given by \cite{Harvey1,CY,Minasian,Dasgupta,Stefanski,SS}
\begin{eqnarray}  S_{4D7} + S_{O7} & \supset &  - \int_{\Sigma^8} G_5 \wedge (\mu_{D7} Y(D7)_{3}^{(0)} + \mu_{O7} Y(O7)_{3}^{(0)} ). \label{WZaction} \end{eqnarray}
We are using standard anomaly descent notation, and we work in units where $4\pi^2 \alpha' = 1$, so that $\mu_{D7} = 2\pi$, $\mu_{O7} = -4 \mu_{D7}$, and

\begin{eqnarray} Y(D7) & = & ch(F) \wedge \sqrt{ \frac{ \hat{A}(R_{T\Sigma}) }{ \hat{A}(R_{N\Sigma}) }} \ , \\
Y(O7) & = & \sqrt{ \frac{ \hat{L}(R_{T\Sigma}/4) }{ \hat{L}(R_{N\Sigma}/4) }} \ . \end{eqnarray}
Note, however, that we do not put the factor of $1/2$ out in front of \eqref{WZaction}, as in \cite{CY} and \cite{SS}, for instance.  This factor of $1/2$ is appropriate for the anomaly inflow analysis because one deals there with a manifestly electric/magnetic symmetric action.  Here, on the other hand, we only consider $C_0,C_2,C_4$ to be the independent R-R forms, with $C_6,C_8$ being obtained via Poincare duality.  The term in $Y(D7)_{3}^{(0)}$ depending on the $D7$-brane gauge field produced the important ``topological mass'' term in our analysis of the defect spectrum in Section \ref{DefectSection}.  Here we focus on the gravitational terms; then $ch(F)$ simply gives\footnote{Or a factor of eight if we are in the orientifold limit, since then the gauge group is $SO(8)$.  However in that case, we use the Type I charge $\mu_{D7}^{(\textrm{I})} = \frac{1}{2} \mu_{D7}^{(\textrm{II})}$.  The net factor is the same in either case.} a factor of four from tracing the identity in the flavor gauge group.  Using
\begin{eqnarray} \hat{A}(R) & = & 1 - \frac{1}{24} p_1(R) + \cdots \ , \\
\hat{L}(R/4) & = & 1 + \frac{1}{48} p_1(R) + \cdots \ , \end{eqnarray}
we have
\begin{eqnarray} S_{4D7} + S_{O7} & \supset &  - 8\pi \int G_5 \wedge \displaystyle\biggl( - \frac{1}{32} p_1(R_{T\Sigma^8}) + \frac{1}{32} p_1(R_{N\Sigma^8}) \displaystyle\biggr)_{3}^{(0)} . \label{wz} \end{eqnarray}

Now let us evaluate this term in the supergravity background of Section \ref{GravitySection}, but with the $SO(6)$ gauge field turned on \eqref{gaugemetric}.  We'll want to use $G_5$ to integrate over $S^5$, so all three legs of $p_1(R)_{3}^{(0)}$ should be taken along $AdS_3$.  In this case we can ignore the normal bundle terms since the pullback of them to the brane will induce couplings to the $D7$-brane scalar.  Restricting $p_1(R_{T\Sigma})_{3}^{(0)}$ to terms that involve only legs along $AdS_3$, one finds
\begin{eqnarray} p_1(R_{T\Sigma})_{3}^{(0)} \displaystyle\bigg|_{AdS} = - \frac{1}{8\pi^2} \textrm{tr}(R^2)_{3}^{(0)} \displaystyle\bigg|_{AdS} = -\frac{1}{8\pi^2} (\tilde{\omega}_3(R_{AdS}) + \tilde{\omega}_3(\mathcal{A})). \label{res1} \end{eqnarray}
The $\tilde{\omega}_3(R_{AdS})$ term will contribute to the gravitational anomaly, but we aren't interested in this now.  Note that the trace in $\tilde{\omega}_3(\mathcal{A})$ is being taken in the vector of $SO(6)$ (which is why we are using the tilde):
\begin{eqnarray} \tilde{\omega}_3(\mathcal{A}) = \textrm{tr}_{\mathbf{6}}(\mathcal{A} \wedge d \mathcal{A} + \frac{2}{3} \mathcal{A}^3). \end{eqnarray}

We will want to convert this to a trace in the $\mathbf{4}$.  For an $SO(6)$ bundle over a base, one has the relation
\begin{eqnarray} ch_2S(N) = p_1(N) \end{eqnarray}
between the second Chern class of the spinor bundle and the first Pontrjagin class.  This implies that
\begin{eqnarray} \textrm{tr}_{\mathbf{4} \oplus \bar{\mathbf{4}}}(R^2) = \textrm{tr}_{\mathbf{6}}(R^2), \end{eqnarray}
where $R$ is the curvature of the bundle.  (Recall that we are identifying $\mathcal{A}$ with the spin connection, $\omega$, on the bundle).  Furthermore, for $SU(4)$ one has
\begin{eqnarray} \textrm{tr}_{\mathbf{4} \oplus \bar{\mathbf{4}}}(R^2) = 2 \textrm{tr}_{\mathbf{4}}(R^2). \end{eqnarray}
Putting all of this together we find that
\begin{eqnarray} p_1(R_{T\Sigma})_{3}^{(0)} \displaystyle\bigg|_{AdS} \supset - \frac{1}{8\pi^2} 2 \textrm{tr}_{\mathbf{4}}(\mathcal{A} \wedge d \mathcal{A} + \frac{2}{3} \mathcal{A}^3) \equiv -\frac{1}{4\pi^2} \omega_3(\mathcal{A}), \end{eqnarray}
and thus
\begin{eqnarray} S_{4D7} + S_{O7} & \supset & - \frac{\pi}{4} \cdot \frac{1}{4\pi^2} \int_{AdS_3 \times S^5} G_5 \wedge \omega_3(\mathcal{A}). \label{intresult} \end{eqnarray}

We should also worry about how $G_5$ depends on the background gauge field.  This is discussed in \cite{HansenKraus}, for example.  One must take
\begin{eqnarray} G_5 = Q \pi^3 ( e_5 - \chi_5 ) + \textrm{ self-dual} \end{eqnarray}
where $e_5$ is the global angular form on the sphere bundle, satisfying $\int_{S^5} e_5 = 1$ along every fiber, and $\pi^3 = vol_{S^5}$.  $d\chi_5 = \chi_6$ is the Euler class of the sphere bundle.  The subtraction of $\chi_5$ is required for closure, $dG_5 = 0$, (away from the source).  Now, $\chi_5$ has a nontrivial gauge variation, $\delta \chi_5 = d\chi_4$, but both $\chi_5$ and $\chi_4$ would have all legs along the base $AdS_3$ and in fact must simply vanish.  Thus $G_5$ will not contribute to the gauge variation and we may simply take
\begin{eqnarray} G_5 = (1 - \ast) 16 \pi \alpha'^2 N_c \epsilon_{S^5}  \quad \rightarrow \quad (1- \ast) \frac{N_c}{ \pi^3} \epsilon_{S^5}, \label{res2} \end{eqnarray}
where $\epsilon_{S^5}$ is the volume form on the unit five-sphere.  Here we have converted the standard result for the 5-form charge to units where $4\pi^2 \alpha' = 1$.

Plugging \eqref{res2} into \eqref{intresult} we obtain
\begin{eqnarray} S_{4D7} + S_{O7} & \supset & -\frac{1}{16\pi} \cdot \frac{N_c}{\pi^3} \int_{AdS_3 \times S^5} \epsilon_{S^5} \wedge \omega_3(\mathcal{A}) \nonumber \\
& = & - \frac{N_c}{16 \pi} \int_{AdS_3} \omega_3(\mathcal{A}). \end{eqnarray}
Now recall that this is the result for one set of $4D7+O7$'s.  Multiplying by $4$ for four sets and comparing with \eqref{whatwewant}, we find
\begin{eqnarray} k = -N_c \ . \end{eqnarray}
The negative simply indicates that the number of \emph{right-handed} minus \emph{left-handed} Weyl fermions transforming in the $\mathbf{4}$ of $SO(6)$ is $N_c$.  This is indeed what we found in the field theory.  We feel that this result gives more support to the proposed duality.  However, strictly speaking, it only shows that if the duality is correct, then there should be no such $AdS_3$ Chern-Simons term resulting from the dimensional reduction of the IIB supergravity action on $\Sigma^2 \times S^5$.

\section{Discussion and open questions}   \label{DiscussionSection}

In this paper we have presented evidence for the AdS/dCFT conjecture involving $D3$-branes and 7-branes intersecting transversely in $1+1$-dimensions.  The correspondence for this particular intersection is special for two reasons.  Firstly, the localized modes on the defect in the field theory are chiral.  To our knowledge, defect mode-operator maps for chiral intersections have not been explicitly worked out before.  In Section \ref{DefectSection} we performed the complete K-K reduction of $D7$-brane open string modes propagating on an $AdS_3 \times S^5$ slice of the near-horizon geometry of the $D3$-branes.  We then enumerated a class of defect operators from the field theory that fit into primary multiplets of the supersymmetry algebra, and exhibited a mode-operator map.  Using the $AdS_3/CFT_2$ mass-dimension relations, we found that the proposed correspondence implies a non-renormalization theorem for the operators.  This is quite interesting because, unlike other AdS/dCFT systems previously studied, superconformal symmetry is completely broken here and there is no algebraic argument protecting operator dimensions.

The second reason this system is special is that the 7-branes can not be treated as probes--on either side of the correspondence.  On the field theory side, the low energy theory on the $D3$-branes had to be considered in the supergravity background of the 7-branes.  This led to interesting consequences such as zero-modes of $D3$-brane fields that localize to the intersection.  On the supergravity side, we considered the fully backreacted geometry produced by the $D3$- and 7-branes, $AdS_3 \times_w \Sigma^2 \times S^5$.  The chiral zero-modes cause a global anomaly in the $SU(4)$ R-symmetry current that localizes to the intersection in the field theory.  In section \ref{RSymmetrySection} we found the supergravity dual signature of this--a term in the effective action involving the $SO(6)$ gauge field, whose gauge variation localized to the $AdS_3$ boundary.

The correspondences between defect modes and operators  worked out here and in
\cite{DeWolfe,Erdmenger} are fascinating because they include a sector relating operators at the brane intersection to modes of the DBI action in a curved background. This sector thus involves an open string-open string duality and might be more amenable to an explicit proof of the duality.

As we discussed in \cite{HR}, this system of $D3$- and 7-branes is T-dual to $N_c$ $D1$-strings in Type I on $T^2$.  As in the $D3$- and 7-brane picture, this theory is not conformal.  In the IR the coupling constant becomes strong and there is conjectured to be a superconformal fixed point corresponding to the Heterotic matrix string theory \cite{BSS,BM,Lowe,Rey}.

Going to the IR of the field theory corresponds to probing the interior of the supergravity solution.  Thus one might have hoped that by zooming in on the intersection region of the supergravity solution \eqref{globalmetric}, one would find a conformal enhancement of supersymmetry and a holographic dual description of the heterotic string.  We have shown, however, that the supergravity solution does not have any such enhancement.  This apparent contradiction is resolved by computing the following curvature invariant:
\begin{eqnarray} R_{MNPQ} R^{MNPQ} & = & 16 \lambda^{-4} e^{-2 a} (\partial_z \partial_{\bar{z}} a)^2 \frac{R^4}{r^4} + 16 \lambda^{-2} e^{-a} (\partial_z \partial_{\bar{z}} a) \frac{1}{r^2} + O(r^0) .  \end{eqnarray}
Thus we see that near the intersection, where $\partial_z a$ is large and $r \rightarrow 0$, the curvature is blowing up.  Thus, as with other supergravity descriptions of the heterotic string \cite{Lapan,Dabholkar,Johnson,Kraus}, $\alpha'$ corrections will need to be taken into account and the geometry will be modified. One might expect the modifications to produce an asymptotic $AdS_3$
factor which would be dual to the IR conformal fixed point that is expected to exist for the field theory.
It would be interesting to explore this further.

Finally, our results suggest that that there should be a non-renormalization theorem given
the exact matching we find between weak and strong coupling in spite of the lack of conformal
invariance. This also deserves further investigation.

\vspace{.3cm}

\section*{Acknowledgements}
We thank O. Lunin and J.  McOrist for helpful conversations.
The work of JH and AR  was supported in part by NSF
Grant No. PHY-00506630 and NSF Grant 0529954. Any opinions,
findings, and conclusions or recommendations expressed in this
material are those of the authors and do not necessarily reflect the
views of the National Science Foundation. AR acknowledges support
from GAANN grant P200A060226.

\appendix

\section{Supersymmetries of the supergravity background} \label{PreservedSUSY}

In this section we construct the supersymmetries preserved by the global geometry presented in Section \ref{GravitySection}.  We show that there is no enhancement of supersymmetry as one goes to the near-horizon region of the $D3$-branes.

We require that the supersymmetry variations of the dilatino and gravitino\footnote{The extra factor of $1/4$ in front of the $G^{(5)}$ term may appear nonstandard.  We follow the conventions of \cite{JoePoeII} for the normalization of the R-R form, which differs from the original work of Schwarz \cite{SchwarzIIB} by a factor of four.  The difference can be seen by comparing the IIB Einstein equation that results from varying the pseudo-action of \cite{JoePoeII} to the Einsein equation in \cite{SchwarzIIB}.  The conventions we chose were fixed by our normalization of the charge in the five-form flux \eqref{RR5form}, which is standard in AdS/CFT.} vanish in the background,
\begin{eqnarray}  \delta \lambda  & = & -i \frac{ \Gamma^M \partial_{M} \tau}{(\tau - \bar{\tau})} \mathcal{B}^\ast \epsilon^\ast = 0 \ , \\
\delta \psi_M & = & \displaystyle\biggl( \nabla_M - \frac{ \partial_M (\tau + \bar{\tau})}{ 4( \tau - \bar{\tau}) } \displaystyle\biggr) \epsilon + \frac{i}{4 \cdot 480} (\Gamma \cdot G^{(5)}) \Gamma_M \epsilon = 0 \ , \end{eqnarray}
and look for solutions to these equations for the SUSY parameter $\epsilon$.  This spinor is complex Weyl, given by $\epsilon = \epsilon_1 + i \epsilon_2$ where $\epsilon_{1,2}$ are Majorana-Weyl.  Since the IIB supercharges are right-handed in our conventions, $\epsilon$ will be left-handed--this is so that $\epsilon Q$ is a Lorentz scalar.  It follows that the gravitino, $\psi_M$, is left-handed, while the dilatino, $\lambda$, is right-handed.  The covariant derivative in the gravitino variation may be written as $\nabla_M - \frac{i}{2} Q_M$, with $Q_M$ the $U(1)$ connection as in \eqref{U1connection}.  The sign of the $U(1)$ charge of $\epsilon$ is also dependent on the chirality choice for the IIB supercharges.  $\mathcal{B}$ is the $d=10$ charge conjugation matrix, satisfying $\mathcal{B} \Gamma^M \mathcal{B}^\ast = (\Gamma^{M})^\ast$, $\mathcal{B} \mathcal{B}^\ast = 1$.

\subsection{Spin connection, gamma matrices, and ansatz for $\epsilon$}

Spacetime coordinates are $x^M = (x^\mu,z,\bar{z},r, \theta^\alpha)$; we will sometimes use $x^m = (x^\mu, r)$ for the $AdS_3$ coordinates.  Corresponding tangent space directions are underlined.  The nonzero components of the vielbein associated with the metric \eqref{globalmetric} are
\begin{eqnarray}  \begin{array}{c} e^{\underline{\mu}}_{\phantom{\underline{\mu}} \nu} = f^{-1/4} \delta_{\nu}^{\underline{\mu}} \ , \\  e^{\underline{r}}_{\phantom{\underline{r}} r} = f^{1/4} \ , \end{array}   \qquad  \begin{array}{c} e^{\underline{z}}_{\phantom{\underline{z}} z} = e^{\underline{\bar{z}}}_{\phantom{\underline{\bar{z}}} \bar{z}} = \lambda e^{a/2} f^{-1/4} \ , \\ e^{\underline{\alpha}}_{\phantom{\underline{\alpha}} \beta} = r f^{1/4} e^{(unit) \underline{\alpha}}_{\phantom{(unit) \underline{\alpha}} \beta} \ . \end{array} \end{eqnarray}
Note that we take the flat metric in the $z,\bar{z}$ directions to be $\eta_{\underline{zz}} = \eta_{\underline{\bar{z}\bar{z}}} = 0$ and $\eta_{\underline{z \bar{z}}} = \eta_{\underline{\bar{z} z}} = 1/2$.  We find the nonzero components of the spin connection to be
\begin{eqnarray} \omega_{\underline{\nu} \underline{r}, \mu} =  \frac{R^4}{r^5 f^{5/4}} e_{\underline{\nu} \mu} \ , \qquad \omega_{\underline{\bar{z}} \underline{r}, z} = \frac{R^4}{r^5 f^{5/4}} e_{\underline{\bar{z}} z} \ , \qquad \omega_{\underline{z} \underline{r}, \bar{z}} = \frac{R^4}{r^5 f^{5/4}} e_{\underline{z} \bar{z}} \ , \nonumber \end{eqnarray}
\begin{eqnarray} \omega_{\underline{\bar{z}} \underline{z}, z} = \frac{1}{4} \partial_z a \ ,  \qquad \omega_{\underline{z} \underline{\bar{z}}, \bar{z}} = \frac{1}{4} \partial_{\bar{z}} a \ , \nonumber \end{eqnarray}
\begin{eqnarray} \omega_{\underline{\beta} \underline{r}, \alpha} = - \frac{1}{f} e^{(unit)}_{\underline{\beta} \alpha} \ , \qquad \omega_{\underline{\beta} \underline{\gamma}, \alpha} = \omega_{\underline{\beta} \underline{\gamma}, \alpha}^{(unit)} \ . \label{spinconnection} \end{eqnarray}
As we take the near horizon limit,
\begin{eqnarray} \lim_{r \rightarrow 0} \frac{R^4}{r^5 f^{5/4}} = \frac{1}{R} \displaystyle\biggl(1 + O(r/R)^4 \displaystyle\biggr), \qquad \lim_{r \rightarrow 0} \frac{1}{f} = O(r/R)^4. \end{eqnarray}
In particular, $\omega_{\underline{\beta} \underline{r}, \alpha} \rightarrow 0$ while $\omega_{\underline{m}\underline{n}, p}$ goes to the standard $AdS_3$ result.

We take the following basis for the $SO(1,9)$ gamma matrices:
\begin{eqnarray} \Gamma^{\underline{m}} & = & \sigma^1 \otimes \mathbbm{1}_2 \otimes \tau^{\underline{m}} \otimes \mathbbm{1}_4 \ , \qquad (m = 0,1,r), \\
\Gamma^{\underline{1+i}} & = & \sigma^3 \otimes \sigma^{i} \otimes \mathbbm{1}_2 \otimes \mathbbm{1}_4 \ , \qquad (i = 1,2), \\
\Gamma^{\underline{4+\alpha}} & = & -\sigma^2 \otimes \mathbbm{1}_2 \otimes \mathbbm{1}_2 \otimes \rho^{\underline{\alpha}} \ , \qquad (\alpha = 1,\ldots 5), \end{eqnarray}
where the $\sigma^i$ are Pauli matrices.  The $\tau^{\underline{m}}$ are the gamma matrices for $SO(1,2)$, which we will think of as the structure group of the tangent bundle of the near horizon $AdS_3$.  They may be taken as real; for instance, $\tau^{\underline{0}} = i \sigma^2, \tau^{\underline{1}} = \sigma^1, \tau^{\underline{r}} = \sigma^3$.  The $\sigma^i$, for $i=1,2$, are gamma matrices for the $SO(2)$ structure group of $T\Sigma^2$, and the $\rho^{\underline{\alpha}}$ are gamma matrices for the $SO(5)$ structure group of $TS^5$.  They can not be taken purely real or purely imaginary.  We can have, say, $\rho^{\underline{1}},\rho^{\underline{3}}$ real and $\rho^{\underline{2}},\rho^{\underline{4}}$ imaginary.

The $10$-dimensional ``$\gamma^5$'' is given by
\begin{eqnarray} \bar{\Gamma} & = & \prod \Gamma^{\underline{M}} = \left( \begin{array}{c c} - \sigma^3 \otimes \mathbbm{1}_8 & 0 \\ 0 & \sigma^3 \otimes \mathbbm{1}_8 \end{array}\right), \end{eqnarray}
so that
\begin{eqnarray} L_{(10)} = \frac{1}{2} (1 + \bar{\Gamma}) = \left( \begin{array}{c c} R_{(2)} & 0 \\ 0 & L_{(2)} \end{array} \right) \otimes \mathbbm{1}_8 \ , \end{eqnarray}
where $L_{(2)},R_{(2)}$ are the Weyl projectors for $SO(2)$.  Since $AdS_3$ and $S^5$ are maximally symmetric spaces, the Killing spinors form a complete basis for general $SO(1,2)$ and $SO(5)$ spinors.  Hence we make the following ansatz for $\epsilon$:
\begin{eqnarray} \epsilon = \left( \begin{array}{c} 0 \\ \zeta_{-}^{a I} (r,z,\bar{z}) \\ \zeta_{+}^{a I} (r,z,\bar{z}) \\ 0 \end{array}\right) \otimes \chi_a \otimes \eta_I \qquad (\textrm{sum over } a,I). \end{eqnarray}
Here $\zeta_{\pm}$ are one-component Weyl spinors of $SO(2)$, $I = 1,\ldots,4$ labels the four Killing spinors of $S^5$, while $a = l,l_\infty$ labels the 2 Killing spinors of $AdS_3$.  They satisfy
\begin{eqnarray}  \nabla_{m} \chi_a = \frac{1}{2 R} \tau_m \chi_a \ , \qquad \nabla_{\alpha}^{(unit)} \eta_I =  \pm \frac{i}{2} \rho_{\alpha}^{(unit)} \eta_I \ . \end{eqnarray}
The two sign choices for $\eta$ are related by charge conjugation, and ``unit'' refers to the unit 5-sphere.  The labeling for the AdS killing spinors comes from the following.  One Killing spinor is an eigenvector of $\tau^{\underline{r}}$ and the other is only an eigenvector in the limits $r \rightarrow 0$ or $r \rightarrow \infty$:
\begin{eqnarray} \tau^{\underline{r}} \chi_l = \chi_l \ , \qquad \begin{array}{c c c}  \lim_{r \rightarrow 0}  \tau^{\underline{r}} \chi_{l_\infty} & = & - \chi_{l_\infty} \\ \lim_{r \rightarrow \infty}  \tau^{\underline{r}} \chi_{l_\infty} & = & \chi_{l_\infty} \end{array}. \end{eqnarray}
The spinors $\zeta_{\pm}^{aI}$ should only depend on $r,z,\bar{z}$ in order to respect the $SO(1,1) \times SO(6)$ isometry.

\subsection{The dilatino equation}

Let us rewrite $\delta \lambda = 0$ slightly.  Taking the conjugate and then acting on the left with $\mathcal{B}^\ast$ gives
\begin{eqnarray} -i \frac{ \mathcal{B}^\ast (\Gamma^M)^\ast \mathcal{B} \partial_M \bar{\tau} }{\tau - \bar{\tau}} \epsilon = 0 \quad \Rightarrow \quad \Gamma^M \partial_M \bar{\tau} \epsilon = 0 \ . \end{eqnarray}
Using the fact that $\bar{\tau} = \bar{\tau}(z)$, we conclude that
\begin{eqnarray} \frac{1}{\lambda} \partial_z \bar{\tau}(z) e^{-a/2} f(r)^{1/4} \Gamma^{\underline{z}} \epsilon = 0 \ , \end{eqnarray}
where $\Gamma^{\underline{z}} = \Gamma^{\underline{2}} + i \Gamma^{\underline{3}}$.  The factor in front of $\Gamma^{\underline{z}} \epsilon$ does not vanish as we go near either the $D3$-branes or the 7-branes (or their intersection).  Hence we must strictly require that
\begin{eqnarray} \Gamma^{\underline{z}} \epsilon = 0 \ , \end{eqnarray}
regardless of any sort of near-horizon limit we may consider.  Given the explicit form of the gamma matrices, one easily finds that this implies $\zeta_{-}^{aI} = 0$, and thus
\begin{eqnarray} \epsilon = \left( \begin{array}{c} 0 \\ 0 \\ \zeta_{+}^{a I} (r,z,\bar{z}) \\ 0 \end{array}\right) \otimes \chi_a \otimes \eta_I \ . \end{eqnarray}
This is just the usual projection for the 7-brane supergravity solution.

\subsection{The gravitino equations}

We will evaluate the equation for $M = \mu,r,z,\bar{z},\alpha$ separately.  First note that
\begin{eqnarray} \Gamma \cdot G^{(5)} & = & 5! \frac{4 R^4}{r^5 f^{5/4}} ( \Gamma^{\underline{01234}} + \Gamma^{\underline{56789}} ) = 240 i \frac{4 R^4}{r^5 f^{5/4}} \left( \begin{array}{c c c c} 0 & 0 & 0 & 0 \\ 0 & 0 & 0 & -1 \\ 1 & 0 & 0 & 0 \\ 0 & 0 & 0 & 0 \end{array} \right) \otimes \mathbbm{1}_8 \ . \label{GammadotG} \end{eqnarray}

\emph{The $M=\mu,r$ equations.}  We find these to be
\begin{eqnarray} \zeta_{+}^{aI} \displaystyle\biggl[ \partial_\mu + \frac{R^4}{2 r^5 f^{5/4}} e_{\underline{\nu} \mu} \tau^{\underline{\nu} \underline{r}} - \frac{R^4}{2 r^5 f^{5/4}} e_{\underline{\nu} \mu} \tau^{\underline{\nu}} \displaystyle\biggr] \chi_a & = & 0 \ , \\
(\partial_r \zeta_{+}^{aI} ) \chi_a + \zeta_{+}^{aI} \displaystyle\biggl[ \partial_r - \frac{R^4}{2 r^5 f^{5/4}} e_{\underline{r} r} \tau^{\underline{r}} \displaystyle\biggr] \chi_a & = & 0 \ . \end{eqnarray}
Using the properties $\tau^{\underline{r}} \chi_l = \chi_l$ and $\chi_l = \chi_l(r)$, one sees that the $\mu$-equation is satisfied for $\chi_l$.  The $r$-dependence of $\zeta_{+}^{lI}$ can also be determined such that the $r$-equation is satisfied.  As $r/R \rightarrow 0$, $\zeta_{+}^{lI} \rightarrow 1$ and $\partial_r \zeta_{+}^{lI} \rightarrow 0$.  For $\chi_{l_\infty}$ on the other hand, one is forced to set $\zeta_{+}^{l_\infty I} = 0$.  However, in the near-horizon limit the equations become
\begin{eqnarray}  \displaystyle\biggl[ \nabla_{m}^{(AdS)} - \frac{1}{2 R} \tau_{m}^{(AdS)} \displaystyle\biggr] ( \zeta_{+}^{aI} \chi_a) = 0 \qquad (\textrm{near horizon}), \end{eqnarray}
which are satisfied for both spinors by taking $\zeta_{+}^{aI}$ constant.

\emph{The $S^5$ equations.}  We find
\begin{eqnarray} \zeta_{+}^{aI} \displaystyle\biggl[ \chi_a \otimes \displaystyle\biggl( \nabla_{\alpha}^{(unit)} - \frac{i R^4}{r^4 f} \rho_{\alpha}^{(unit)} \displaystyle\biggr) \eta_I - \tau^{\underline{r}} \chi_a \otimes \frac{i}{2 f} \rho_{\alpha}^{(unit)} \eta_I \displaystyle\biggr] = 0 \ . \end{eqnarray}
For $\chi_l$, the second and third terms can be added to give precisely the coefficient $i/2$.  Hence the equation is \emph{always satisfied} for $\zeta_{+}^{lI}$, due to the fact that $\eta_I$ is a Killing spinor.  We must set $\zeta_{+}^{l_\infty I} = 0$ for the general background.  However, in the near horizon limit, the last term vanishes while the first two become
\begin{eqnarray} \zeta_{+}^{aI} \chi_a \otimes \displaystyle\biggl( \nabla_{\alpha}^{(unit)} - \frac{i}{2} \rho_{\alpha}^{(unit)} \displaystyle\biggr) \eta_I = 0 \qquad (\textrm{near horizon}), \end{eqnarray}
so that this equation is also consistent with enhanced solutions.

\emph{The $M = z,\bar{z}$ equations.}  The $M = z$ equation leads to two equations, due to the fact that $\omega_{\underline{\bar{z}}\underline{r}, z} \Gamma^{\underline{\bar{z}} \underline{r}}$ in the covariant derivative and $(\Gamma \cdot G^{(5)}) \Gamma_z$ both act as off-diagonal matrices on the column vector $(0,0,\zeta,0)^T$.  We find
\begin{eqnarray}  \displaystyle\biggl( \partial_z - \frac{1}{4} \partial_z a + \frac{1}{4} \partial_{z} \log{\tau_2} \displaystyle\biggr) \zeta_{+}^{aI} & = & 0 \ , \\
\frac{R^4}{r^5 f^{5/4}} e_{\underline{\bar{z}} z} \zeta_{+}^{aI} ( \tau^{\underline{r}} \chi_a - \chi_a ) & = & 0 \ . \end{eqnarray}
For the $M= \bar{z}$ equation, both of the off-diagonal terms that produced the extra equation above vanish when acting on $\epsilon$, and we just have
\begin{eqnarray} \displaystyle\biggl( \partial_{\bar{z}} + \frac{1}{4} \partial_{\bar{z}} a - \frac{1}{4} \partial_{\bar{z}} \log{\tau_2} \displaystyle\biggr) \zeta_{+}^{aI} & = & 0 \ . \end{eqnarray}
This equation combined with the first of the $z$ equations fixes the $z,\bar{z}$ dependence of $\zeta_{+}^{aI}$.  The solution has the form
\begin{eqnarray} \zeta_{+}^{aI}(r,z,\bar{z}) = A(r) \displaystyle\biggl( \frac{g(z)}{\bar{g}(\bar{z})} \displaystyle\biggr)^{1/4}, \label{zetasol} \end{eqnarray}
where $g(z)$ is the holomorphic function appearing in the metric.  This $z,\bar{z}$ dependence is the standard result for the killing spinors of the 7-brane supergravity solution.

The second $z$ equation, on the other hand, is clearly only solved for $\zeta_{+}^{lI}$.  Furthermore, in the near horizon limit it becomes
\begin{eqnarray}  \frac{r}{2 R^2} \lambda e^{a/2} \zeta_{+}^{aI} ( \tau^{\underline{r}} \chi_a - \chi_a ) = 0 \ . \end{eqnarray}
In all of the other equations, the constraint that $\zeta_{+}^{l_\infty I} = 0$ vanished like $O(r/R)^4$, whereas here it only vanishes like $O(r/R)$.  Hence, while the $AdS_3$ and $S^5$ equations are consistent with the enhancement of supersymmetry in the near horizon, the $\Sigma^2$ equations are not.  We conclude that there is no enhancement.

The supersymmetries preserved in the global solution are of the form
\begin{eqnarray} \epsilon = \left(\begin{array}{c} 0 \\ 0 \\ \zeta_{+}^{lI}(r,z,\bar{z}) \\ 0 \end{array}\right) \otimes \chi_l \otimes \eta_I \ , \end{eqnarray}
where $\zeta_{+}^{lI}$ is given by \eqref{zetasol} and $\chi_l,\eta_I$ satisfy
\begin{eqnarray} \begin{array}{r c l} \nabla_{m}^{(AdS)} \chi_l & = & \frac{1}{2 R} \tau_{m}^{(AdS)} \chi_l \\  \tau^{\underline{r}} \chi_l & = & \chi_l \end{array} , \qquad  \nabla_{\alpha}^{(unit)} \eta_I = \frac{i}{2} \rho_{\alpha}^{(unit)} \eta_I \ . \end{eqnarray}
This corresponds to 4 complex, or 8 real supercharges.  Given the identification of $\tau^{\underline{r}}$ with the ``$\gamma^5$'' of the $1+1$-dimensional boundary and the isometries of $S^5$ with an $SO(6)$ R-symmetry group, we conclude that $\epsilon$ transforms in the $(1/2, \mathbf{4})_+$ of $SO(1,1) \times SO(6) \times SO(2)$.  This implies that the preserved supercharges, $Q$, transform in the $(-1/2, \bar{\mathbf{4}})_{-}$, which is consistent with our analysis of the field theory side \cite{HR}.

\section{The $D7$-brane fermion action} \label{D7fermion}

We begin with the action
\begin{eqnarray} S_{D7}^{ferm.} = \frac{i T_{D7}}{2} \int d^8 \xi e^{\phi} \sqrt{-g} \textrm{tr} \displaystyle\biggl( \bar{y} (1 - \tilde{\Gamma}_{D7}) (e^{-\phi/4} \hat{\Gamma}^m \breve{D}_m - \breve{\Delta} ) y \displaystyle\biggr). \label{MMSaction} \end{eqnarray}
Let us review what these various quantities are, following \cite{MMS}.  In this section we denote by $m,n,\ldots$ and $a,b,\ldots$ worldvolume/tangent space indices along the brane, and $M,N,\ldots$, $A,B,\ldots$ denote spacetime/tangent space indices in the bulk.  We also use $i,j,\ldots$ and $\underline{i},\underline{j},\ldots$ to denote spacetime and tangent space indices transverse to the brane.  We have
\begin{eqnarray} y = \left(\begin{array}{c} y_1 \\ y_2 \end{array}\right), \end{eqnarray}
where $y_1,y_2$ are each 32-component $d=10$ spinors, satisfying Majorana and Weyl constraints, both of the same chirality.  Also,
\begin{eqnarray} \hat{\Gamma}^A = \left( \begin{array}{c c} \Gamma^A & 0 \\ 0 & \Gamma^A \end{array}\right) = I_2 \otimes \Gamma^A \ , \qquad \hat{\bar{\Gamma}} = \left(\begin{array}{c c} \bar{\Gamma} & 0 \\ 0 & - \bar{\Gamma} \end{array}\right) = \sigma^3 \otimes \bar{\Gamma} \ , \end{eqnarray}
where $\Gamma^A$ are the $d=10$ gamma matrices and $\bar{\Gamma}$ the generalized ``$\gamma^5$.''  $\tilde{\Gamma}_{D7}$ is given by
\begin{eqnarray} \tilde{\Gamma}_{D7} = -i \sigma^2 \otimes \frac{1}{8! \sqrt{-g}} \varepsilon^{m_1 \cdots m_8} \Gamma_{m_1 \cdots m_8} \ , \end{eqnarray}
where $\varepsilon^{0\cdots 7} = 1$ and $\Gamma_{m_1 \cdots m_8} = \Gamma_{[m_1} \cdots \Gamma_{m_8]}$.  $\frac{1}{2}(1 - \tilde{\Gamma}_{D7})$ is a kappa symmetry projection operator that will remove half of the degrees of freedom.  Finally, $\breve{D}_m$ and $\breve{\Delta}$ are given by
\begin{eqnarray} \breve{D}_m = \mathbbm{1}_2 \otimes \hat{D}^{(0)}_m + \sigma^1 \otimes \hat{W}_m \ , \qquad \breve{\Delta} = \mathbbm{1}_2 \otimes \hat{\Delta}^{(1)} + \sigma^1 \otimes \hat{\Delta}^{(2)} \ , \end{eqnarray}
where
\begin{eqnarray} \hat{D}_{(1,2)M}^{(0)} & = & \partial_M + \frac{1}{4} \displaystyle\biggl( \omega_{AB,M} + \frac{1}{4} \tau_{AB,M} \displaystyle\biggr) \Gamma^{AB} \equiv \tilde{D}_M \ , \label{codiv} \\
\hat{W}_{(1,2)M} & = & \frac{1}{8} \displaystyle\biggl( \mp e^{3 \phi/4} G^{(1)}_{A} \Gamma^A \mp \frac{e^{-\phi/4}}{2 \cdot 5!} G_{ABCDE}^{(5)} \Gamma^{ABCDE} \displaystyle\biggr) e^{ \phi/4} \Gamma_M \ , \\
\hat{\Delta}_{(1,2)}^{(1)} & = & \frac{1}{2} e^{- \phi/4} \Gamma^M \partial_M \phi \ , \\
\hat{\Delta}_{(1,2)}^{(2)} & = & \pm \frac{1}{2} e^{3 \phi/4} G^{(1)}_{A} \Gamma^A \ . \end{eqnarray}
The subscript $(1,2)$ is correlated with the sign.  If the operator acts on $y_1$ the top sign is chosen, if it acts on $y_2$ the bottom sign is chosen.

One important step has already been taken relative to the formulae presented in \cite{MMS}.  Their results are given in string frame and we have converted to Einstein frame\footnote{Note, however, that the R-R forms in \cite{MMS} were already in Einstein frame.} using $g_{MN}^{(s)} = e^{\phi/2} g_{MN}^{(e)}$.  In terms of the vielbeins and inverse vielbeins, $e^{(s)A}_{\phantom{(s)A}M} = e^{\phi/4} e^{(e)A}_{\phantom{(s)A}M}$ and $E_{A}^{(s)M} = e^{-\phi/4} E_{A}^{(e)M}$.  Thus, for instance, one has $\Gamma^{(s)M} = e^{-\phi/4} \Gamma^{(e)M}$ and $G_{A}^{(1)(s)} = e^{-\phi/4} G_{A}^{(1)(e)}$.  The projector $\tilde{\Gamma}_{D7}$ is unchanged; the transformation of the vielbeins in $\Gamma_{m_1 \cdots m_8}$ cancels the factor coming from $\sqrt{-g}$ in the denominator.  Finally, it can be shown that the spin connection gets modified: $\omega_{AB,M}^{(s)} = \omega_{AB,M}^{(e)} + \frac{1}{4} \tau_{AB,M}^{(e)}$, where $\tau_{AB,M}$ is defined through
\begin{eqnarray} \label{taudef1}  \tau_{AB,C} & = & \frac{1}{2} ( \chi_{A,BC} + \chi_{B,CA} - \chi_{C,AB} ), \end{eqnarray}
with $\chi_{A,BC} = -\chi_{A,CB}$ given by
\begin{eqnarray} \label{taudef2}  d\phi \wedge e_A & = & \frac{1}{2} \chi_{A,BC} e^B \wedge e^C \ . \end{eqnarray}

Now let us simplify this action.  We have
\begin{eqnarray} \frac{1}{8! \sqrt{-g}} \varepsilon^{m_1 \cdots m_8} \Gamma_{m_1 \cdots m_8} = \frac{e^{a_1}_{\phantom{a_1}m_1} \cdots e^{a_8}_{\phantom{a_8}m_8} }{8! \sqrt{-g}} \varepsilon^{m_1 \cdots m_8} \Gamma_{a_1 \cdots a_8} = \Gamma_{\underline{0}\cdots \underline{7}} = i \bar{\Gamma}_{(8)} \ ,  \end{eqnarray}
where we've defined $\bar{\Gamma}_{(8)} \equiv i \Gamma^{\underline{0} \cdots \underline{7}}$ which anticommutes with all of the $\Gamma^a$ and squares to one.  The dilaton and axion only depend on coordinates transverse to the brane, so $\Gamma^M \partial_M \phi = \Gamma^i \partial_i \phi$ and $\Gamma^M G_{M}^{(1)} = \Gamma^i G_{i}^{(1)}$.  In particular, when we compute $\Gamma^M \hat{W}_M$, we use $\Gamma^m \Gamma^i \Gamma_m = - 8 \Gamma^i$.  Using the near-horizon form of \eqref{GammadotG} we find
\begin{eqnarray}  G_{ABCDE}^{(5)} \Gamma^m \Gamma^{ABCDE} \Gamma_m & = & \frac{8 \cdot 5!}{R} ( - \Gamma^{\underline{01234}} + \Gamma^{\underline{567689}} ). \end{eqnarray}

Finally, consider the covariant derivative $\tilde{D}_m$, given in \eqref{codiv}.  From the expression \eqref{spinconnection} for the spin connection $\omega_{AB,M}$, we see that when $M$ is an $AdS_3$ or $S^5$ coordinate, so are $A,B$.  Hence $\omega_{AB,m} \Gamma^{AB} = \omega_{ab,m} \Gamma^{ab}$.  Using the definitions \eqref{taudef1},\eqref{taudef2}, we find $\Gamma^m \tau_{AB,m} \Gamma^{AB} = 16 \Gamma^i \partial_i \phi$.  Therefore
\begin{eqnarray} \Gamma^m \tilde{D}_m & = & \Gamma^m \nabla_m + \Gamma^i \partial_i \phi \ , \end{eqnarray}
where $\nabla_m$ is the standard covariant derivative on the brane worldvolume.

Putting these results together, one obtains the following form for the Dirac operator in \eqref{MMSaction}:
\begin{eqnarray} (1 - \tilde{\Gamma}_{D7}) (e^{\frac{-\phi}{4}} \hat{\Gamma}^m \breve{D}_m - \breve{\Delta} ) &=& e^{\frac{-\phi}{4}} \left( \begin{array}{c c} \mathcal{A} + \mathcal{C}^+ & (\mathcal{A} - \mathcal{C}^-) (-i \bar{\Gamma}_{(8)} ) \\ (\mathcal{A} - \mathcal{C}^-) ( i \bar{\Gamma}_{(8)} ) & \mathcal{A} + \mathcal{C}^+ \end{array}\right) \nonumber \\
& \equiv & \mathcal{M} \ ,  \end{eqnarray}
where
\begin{eqnarray} \mathcal{A} & = & \Gamma^m \nabla_m + \frac{1}{2R} ( - \Gamma^{\underline{01234}} + \Gamma^{\underline{56789}} ) (i \bar{\Gamma}_{(8)} ) \ , \\
\mathcal{C}^{\pm} & = & \frac{1}{2} \Gamma^i [\pm \partial_i \phi + e^\phi G_{i}^{(1)} (i \bar{\Gamma}_{(8)} ) ] . \end{eqnarray}
Note that $\mathcal{A}$ anticommutes with $\bar{\Gamma}_{(8)}$ while $\mathcal{C}^{\pm}$ commute with it.  We would like to make a unitary change of variables that diagonalizes $\mathcal{M}$.  Consider $\tilde{y} = U y$ where
\begin{eqnarray} U = \frac{1}{\sqrt{2}} \left( \begin{array}{c c} \bar{\Gamma}_{(8)} & -i \bar{\Gamma}_{(8)} \\ \bar{\Gamma}_{(8)} & i \bar{\Gamma}_{(8)} \end{array}\right). \end{eqnarray}
One can then show that $\bar{y} \mathcal{M} y = \bar{\tilde{y}} \tilde{\mathcal{M}} \tilde{y}$, with $\tilde{M} = - U \tilde{M} U^\dag$ given by
\begin{eqnarray} \tilde{M} & = & 2 e^{-\phi/4} \left(\begin{array}{c c} \mathcal{A} \hat{L}_{(8)} - \mathcal{C}^+ \mathbbm{1} + \mathcal{C}^- \bar{\Gamma}_{(8)} & 0 \\ 0 & \mathcal{A} \hat{R}_{(8)} - \mathcal{C}^+ \mathbbm{1} - \mathcal{C}^- \bar{\Gamma}_{(8)} \end{array}\right),  \end{eqnarray}
where $\hat{L}_{(8)} = \frac{1}{2} (1 + \bar{\Gamma}_{(8)})$ and $\hat{R}_{(8)} = \frac{1}{2} (1 - \bar{\Gamma}_{(8)})$.  The hats are a reminder that these matrices are still $32 \times 32$.  Note that the original $y_{1,2}$ were both left-handed, $L_{(10)} y_{1,2} = y_{1,2}$, and Majorana, $y_{1,2} = C \bar{y}_{1,2}^{T}$, where $C$ is the $d=10$ charge conjugation matrix.  The new $\tilde{y}_{1,2}$ are related to the old by $\tilde{y}_{1} = \frac{1}{\sqrt{2}} \bar{\Gamma}_{(8)} (y_1 - i y_2)$ and $\tilde{y}_2 = \frac{1}{\sqrt{2}} \bar{\Gamma}_{(8)} (y_1 + i y_2)$.  It is easy to show that the $\tilde{y}$ satisfy the following Weyl and Majorana conditions:
\begin{eqnarray} L_{(10)} \tilde{y}_{1,2} = \tilde{y}_{1,2} \ , \qquad \begin{array}{c} \tilde{y}_1 = C \bar{\tilde{y}}_{2}^{T} \\ \tilde{y}_{2} = C \bar{\tilde{y}}_{1}^{T} \end{array} . \label{d10MW} \end{eqnarray}

Now let us dimensionally reduce from $d=10$ to $d=8$.  We take the $\Gamma^A$ to have the form
\begin{eqnarray} \Gamma^a = \mathbbm{1}_2 \otimes \gamma^a \ , \quad a = 0,1,4,\ldots,9 \ , \qquad \Gamma^{2,3} = -\sigma^{1,2} \otimes \bar{\gamma} \ . \label{10to8} \end{eqnarray}
These definitions are consistent with those in the previous section.  The $\gamma^a$ are $SO(1,7)$ gamma matrices and $\bar{\gamma} = i \prod \gamma^a$.  Appropriate definitions of the charge conjugation matrices in $d=10$ and $d=8$ exist such that $C = \sigma^1 \otimes c$, and also $c = c^T$.  Therefore we let
\begin{eqnarray} \tilde{y}_{1,2} = \frac{1}{\sqrt{2}} \left( \begin{array}{c} \Psi_{1,2} \\ \chi_{1,2} \end{array}\right), \end{eqnarray}
where
\begin{eqnarray} \begin{array}{c} L_{(8)} \Psi_{1,2} = \Psi_{1,2} \\ R_{(8)} \chi_{1,2} = \chi_{1,2} \end{array} \qquad \textrm{and} \qquad \begin{array}{c} \Psi_{1,2} = c \bar{\chi}_{2,1}^T \\ \chi_{1,2} = c \bar{\Psi}_{2,1}^T \end{array} . \label{d8MW} \end{eqnarray}
Plugging these expressions in we eventually find
\begin{eqnarray} e^{\phi/4} \bar{\tilde{y}} \tilde{\mathcal{M}} \tilde{y}  & = &  \bar{\Psi}_1 [ \slashed{\nabla} + \frac{i}{2R} (-i \gamma^{\underline{014}} + \gamma^{\underline{5 \cdots 9}} ) \bar{\gamma}] L_{(8)} \Psi_1 + \nonumber \\
& & + \bar{\chi}_2 [ \slashed{\nabla} + \frac{i}{2R} (i \gamma^{\underline{014}} + \gamma^{\underline{5 \cdots 9}} ) \bar{\gamma}] R_{(8)} \chi_2 + \nonumber \\
& & + 2 E_{\underline{z}}^{\phantom{\underline{z}} z} [ - \partial_z \phi \bar{\Psi}_2 \chi_2 + i e^\phi G_{z}^{(1)} \bar{\Psi}_1 \chi_1 ] +  \nonumber \\
& & + 2 E_{\underline{\bar{z}}}^{\phantom{\underline{\bar{z}}} \bar{z}} [ \partial_{\bar{z}} \phi \bar{\chi}_1 \Psi_1 + i e^\phi G_{\bar{z}}^{(1)} \bar{\chi}_2 \Psi_2 ].   \end{eqnarray}

At this point it is clear that $\chi_1$ and $\Psi_2$ (which are charge conjugates of each other) are non-dynamical fields.  They represent the extra degrees of freedom that are supposed to be projected out by the $\kappa$-projection operator $\frac{1}{2} (1 - \tilde{\Gamma}_{D7})$.  However, it appears that, rather than getting projected out, they are serving as Lagrange multipliers.  What is the constraint they are imposing?  From the charge conjugation relations it follows that $\bar{\Psi}_2 \chi_2 = - \bar{\Psi}_1 \chi_1$ and $\bar{\chi}_2 \Psi_2 = - \bar{\chi}_1 \Psi_1$.  Therefore the last two terms may be written as
\begin{eqnarray} 2 E_{\underline{z}}^{\phantom{\underline{z}} z} [ \partial_z \phi + i e^\phi G_{z}^{(1)} ] \bar{\Psi}_1 \chi_1  + 2 E_{\underline{\bar{z}}}^{\phantom{\underline{\bar{z}}} \bar{z}} [ \partial_{\bar{z}} \phi - i e^\phi G_{\bar{z}}^{(1)} ] \bar{\chi}_1 \Psi_1 \ . \end{eqnarray}
Observe that
\begin{eqnarray}   \partial_z \phi + i e^\phi G_{z}^{(1)} & = & - \partial_z \log{\tau_2} + \frac{i}{\tau_2} \partial_z \tau_1 = \frac{i}{\tau_2} \partial_z \tau \ , \\
\partial_{\bar{z}} \phi - i e^\phi G_{\bar{z}}^{(1)} & = & - \partial_{\bar{z}} \log{\tau_2} - \frac{i}{\tau_2} \partial_{\bar{z}} \tau_1 = - \frac{i}{\tau_2} \partial_{\bar{z}} \bar{\tau} \ , \end{eqnarray}
where $\tau = \tau_1 + i \tau_2$ is the axidilaton.  But these are precisely the quantities that must vanish in order for the supergravity background to preserve supersymmetry and satisfy the equations of motion.  Hence the $\kappa$-projection simply requires that we work in a consistent background.  Then these terms are indeed removed from the action.

The terms that remain in the action are those involving only $\Psi_1$, or its conjugate $\chi_2$.  In fact, by setting the $\chi_2$ term equal to its transpose, using the charge conjugation relations \eqref{d8MW}, and carefully bringing the resulting charge conjugation matrices together, one can show that the two terms are equivalent.  The action finally boils down to
\begin{eqnarray} S_{D7}^{ferm.} & = & i T_{D7} \int d^8 \xi e^{3\phi/4} \sqrt{-g} \textrm{tr} \displaystyle\biggl( \bar{\Psi}_1 [ \slashed{\nabla} + \frac{i}{2R} (-i \gamma^{\underline{014}} + \gamma^{\underline{5 \cdots 9}} ) \bar{\gamma}] L_{(8)} \Psi_1 \displaystyle\biggr).  \end{eqnarray}

Now we must rescale the metric as discussed above equation \eqref{Mrescale}.  This sends $\sqrt{-g} \rightarrow R^8 \sqrt{-g}$ and $\slashed{\nabla} \rightarrow \frac{1}{R} \slashed{\nabla}$, producing an overall factor of $R^7$.  In the conventions of \cite{MMS}, $2\pi \alpha' = 1$.  Therefore, $T_{D7} R^4 = N_c/(8\pi^4)$ is the factor we obtained in front of the bosonic action.  Hence it would appear that we have an unwanted factor of $R^3 e^{3\phi/4}$.  We would like to make the rescaling
\begin{eqnarray} \Psi_1 & = & R^{-3/2} e^{-3\phi/8} \Psi \, \label{needtorescale} \end{eqnarray}
and then we would have precisely the result quoted in the text \eqref{FermionAction}.  However, we need a good reason to do so, since $e^{-3\phi/8}$ diverges when evaluated on the brane, and $\alpha'$ in $R$ is being taken to zero.  For this we turn to supersymmetry.

We can obtain the supersymmetry variations of the $D7$-brane fields by dimensionally reducing the $d=10$ SYM relations to eight dimensions.  The $d=10$ variations are
\begin{eqnarray} \delta A_M = -i \bar{\zeta} \Gamma_M \lambda \ , \qquad \delta \lambda = \frac{1}{2} F_{MN} \Gamma^{MN} \zeta \ , \end{eqnarray}
where $\lambda,\zeta$ are Majorana-Weyl.  We use the gamma matrix decomposition \eqref{10to8}, and write $\lambda = (\Psi, \Psi^c)^T$, $\zeta = (\xi, \xi^c )^T$, where $\Psi$ and $\xi$ are left-handed in $d=8$, $\Psi^c = c \bar{\Psi}^T$ and similarly for $\xi^c$.  Here we will restrict to the $U(1)$ case, since it will be sufficient for our purposes.  Carrying out the reduction in a curved background leads one to
\begin{eqnarray} \delta A_m & = & -i (\bar{\xi} \gamma_m \Psi - \bar{\Psi} \gamma_m \xi ), \nonumber \\
\delta A_z & = & i e^{\underline{z}}_{\phantom{\underline{z}} z} \bar{\xi^c} \Psi \ , \qquad \delta A_{\bar{z}} = -i e^{\underline{\bar{z}}}_{\phantom{\underline{\bar{z}}} \bar{z}} \bar{\xi} \Psi^c \ , \nonumber \\
\delta \Psi & = & \frac{1}{2} F_{mn} \gamma^{mn} \xi + 2 E_{\underline{z}}^{\phantom{\underline{z}} z} \slashed{\partial} A_z \xi^c \ . \end{eqnarray}

We expect that the $D7$-brane action in the supergravity background preserves half of these supersymmetries.  It is clear, however, that these would be the variations in string frame, where the dilaton sits out in front of all terms with equal weight.  To determine how SUSY acts in Einstein frame, we must Weyl-rescale the variations according to $G_{MN}^{(s)} = e^{\phi/2} R^2 \bar{G}_{MN}^{(e)}$.  There are factors of vielbeins explicitly, and implicitly in the gamma matrices.  Furthermore, observe that the superalgebra is schematically of the form $\{Q, \bar{Q} \} = \Gamma^M \partial_M$.  Since $\Gamma^M \rightarrow e^{-\phi/4} R \Gamma^M$, it follows that the supercharges must be rescaled: $Q^{(s)} = e^{-\phi/8} R^{-1/2} Q^{(e)}$.  The supercharges appear on the left hand side of the variations according to $\delta \varphi \equiv [ Q \xi + \bar{Q} \bar{\xi} , \varphi ]$.  After plugging these factors in, the variations are brought back to canonical form by the field redefinitions
\begin{eqnarray} A_{M}^{(s)} = A_{M}^{(e)} , \qquad \Psi^{(s)} = R^{-3/2} e^{-3\phi/8} \Psi^{(e)} \ . \end{eqnarray}
There is an overall scaling factor that is not fixed by this argument, but is fixed by the fact that the gauge field action is already well defined in Einstein frame and leads to finite energy fluctuations in the Maldacena limit.  Hence supersymmetry requires the rescaling \eqref{needtorescale}.  Finite energy gauge fluctuations are mapped by SUSY to finite energy fermionic fluctuations in the Einstein variable $\Psi^{(e)} \equiv \Psi$.

\end{document}